\documentclass[12pt]{iopart}
\usepackage[latin1]{inputenc}
\usepackage{iopams}

\begin{document}

\title[Modular Transformations in the Schwinger Model]{Dressed Fermions, Modular Transformations and
Bosonization in the Compactified Schwinger Model}

\author{Micha\"el Fanuel$^1$ and Jan Govaerts$^{1,2,}$\footnote{Fellow of the Stellenbosch Institute
for Advanced Study (STIAS), 7600 Stellenbosch, South Africa}}
\vspace{10pt}
\address{$^1$Centre for Cosmology, Particle Physics and Phenomenology (CP3),\\
Institut de Recherche en Math\'ematique et Physique,\\
Universit\'e catholique de Louvain, Chemin du Cyclotron 2, bte L7.01.01,\\
B-1348, Louvain-la-Neuve, Belgium}
\vspace{10pt}
\address{$^2$International Chair in Mathematical Physics\\
and Applications (ICMPA--UNESCO Chair),\\
University of Abomey--Calavi, 072 B. P. 50, Cotonou, Republic of Benin}
\ead{Michael.Fanuel@uclouvain.be, Jan.Govaerts@uclouvain.be}

\begin{abstract}
The celebrated exactly solvable ``Schwinger'' model, namely massless two-dimensional QED, is revisited.
The solution presented here emphasizes the non-perturbative relevance of the topological sector through large gauge transformations
whose role is made manifest by compactifying space into a circle. Eventually the well-known non-perturbative features and solution
of the model are recovered in the massless case. However the fermion mass term is shown to play a subtle role in order to achieve a
physical quantization that accounts for gauge invariance under both small and large gauge symmetries.
Quantization of the system follows Dirac's approach in an explicitly gauge invariant way that avoids any gauge fixing procedure.
\end{abstract}

\section{Introduction}

The gauge invariance principle is an essential cornerstone to the modern approach towards the unification of the fundamental
quantum interactions. The riches of gauge theories are particularly relevant when considered as quantum dynamical systems.
However until only recently the major framework available for studying an interacting quantized theory has been the perturbative approach,
with supersymmetric Yang-Mills theories and M-theory being recent exceptions for which important non-perturbative insights
and results have been achieved \cite{SW1,SW2}. As the latter progress has shown however, the perturbative approach allows to address
only partly the broad and rich physics expected and known to be at work in gauge theories. Especially in the context of
non-supersymmetric Yang-Mills theories coupled to matter new non-perturbative tools need to be imagined and developed
to completely unravel the dynamics of such gauge theories. These issues also involve the so-called Gribov problems
that arise whenever a gauge fixing procedure is considered for the quantized system. One possibility to circumvent these
issues is by avoiding any gauge fixing procedure altogether. Furthermore, in the presence of nontrivial space(time)
topology, the topological properties both of the space of gauge field configurations and of gauge transformations are
expected to play a fundamental role in the non-perturbative dynamics of these theories, in particular for instance in
the possible mechanisms leading to confinement in the case of QCD \cite{'tHooft:1981ht}.

The Schwinger model \cite{Schwinger:1962tp} is one of the very few, if not the only (non-supersymmetric) theoretical laboratory providing
exact and non-perturbative results for a gauge theory. Classical references analyze this model in the Coulomb gauge \cite{LZ,Manton:1985jm}.
However, in the non-perturbative domain usually any gauge fixing procedure induces so-called Gribov problems.
To avoid these difficulties it is possible to apply in the case of this model a manifestly gauge invariant quantization free of gauge fixing,
but rather by relying on a gauge invariant factorization of the physical degrees of freedom~\cite{BB1,BB2}.
Furthermore, invariance under ``large gauge transformations'' is made explicit; space compactification into a circle makes possible
the factorization of gauge symmetries into ``small'' and ``large'' gauge transformations. Finally, space compactifiction also provides
for a regularization of infrared divergencies, while momentum space then becomes discrete thus entailing a system with a countable
albeit infinite set of degrees of freedom, yet without modifiying the super-renormalisable character of the dynamics in the ultraviolet
regime. In the present work a constructive and operatorial approach is favoured over the functional bosonization
one \cite{Avossevou:2002fr} to make the solution as explicit as possible, with the intent as well to extend similar methods
to higher dimensions for which bosonization methods are less readily available. However a most tantalizing feature that is made manifest
through space compactification is the fact that large gauge transformations imply a connection between large and small distance dynamics
in gauge theories. As a matter of fact this observation applies generally to any Yang-Mills gauge theory coupled to charged matter
over a spacetime of non trivial topology, be it as simple as a flat and compact toroidal geometry.

The outline of the work is as follows. In the next Section, the Hamiltonian formulation of the model is reviewed. Section 3. considers
its canonical quantization in careful detail in the fermionic formulation, by paying due attention in particular to large gauge
transformations, namely the topological modular symmetries of the dynamics, an issue which to the best of the authors' knowledge
is new in the literature as well as the new understanding our approach provides for some of the non-perturbative properties of
the Schwinger model. These physical consequences are addressed in the following Sections 4. to 6.
Some concluding remarks are provided in Section 7., with other useful considerations being detailed also in an Appendix.

\section{Hamiltonian Formulation}

In the expression hereafter for the Lagrangien density of the massive or massless Schwinger model,
a possible choice for the Clifford-Dirac algebra of $\gamma^\mu$ matrices ($\mu=0,1$) is taken to be given by
$\gamma^{0}=\sigma^{1}$ and  $\gamma^{1}=\rmi\sigma^{2}$, the chirality matrix then being
$\gamma_{5}=\gamma^{0}\gamma^{1}=-\sigma^{3}$, while the $\sigma^i$ ($i=1,2,3$) stand of course for the usual Pauli matrices.
Further notational conventions include an implicit choice of units such that $\hbar=c=1$, the Levi-Civita tensor $\epsilon^{\mu\nu}$
being defined by $\epsilon^{01}=+1$, and the Minkowski metric taken to be $\eta_{\mu\nu}={\rm diag}\,(1,-1)$.

The starting point of the analysis is the QED Lagrangian density in its explicitly self-adjoint form,
\begin{eqnarray}
\mathcal{L}=-\frac{1}{4}F_{\mu\nu}F^{\mu\nu}+\frac{1}{2}\rmi\overline{\psi}\gamma^{\mu}(\partial_{\mu}+\rmi eA_{\mu})\psi-\frac{1}{2}\rmi(\overline{\partial_{\mu}+\rmi eA_{\mu})\psi}\gamma^{\mu}\psi-\mu\overline{\psi}\psi,
\label{Ldebut}
\end{eqnarray}
with $\overline{\psi}=\psi^\dagger\gamma^0$,
where $\psi$, $A_{\mu}$, $F_{\mu\nu}=\partial_\mu A_\nu-\partial_\nu A_\mu$ and $\mu\ge 0$ denote the Dirac spinor field,
the gauge field, the field strength tensor, and a fermionic mass term, respectively.
In $D=2$ space-time dimensions, the gauge coupling constant $e$ has dimension $[e]=1$
in units of mass, while the gauge and matter fields have mass dimensions $[A_{\mu}]=0$ and $[\psi]=\frac{1}{2}$.
This theory having a coupling constant of strictly positive mass dimension is perturbatively super-renormalizable.
Infrared divergencies inherent to such a theory are regularized in our case by having compactified space into a circle of circumference $L$,
with the further consequence of a discretization of momentum space implying a countable set of quantum modes for the fields.
Given the cylindrical spacetime topology which breaks the symmetry under Lorentz boosts but not under spacetime translations,
the boundary conditions of the fields in the spatial circular direction are taken to be
\begin{eqnarray}
A_{\mu}(t,x+L)=A_{\mu}(t,x),\qquad
\psi(t,x+L)=\exp(-2\rmi \pi\lambda)\psi(t,x),
\end{eqnarray}
where $\lambda\in [0,1[$ is a fermionic holonomy parameter.

The $U(1)$ gauge symmetry of the model acts through the transformations
$A_{\mu}'(t,x)=A_{\mu}(t,x)+\frac{1}{e}\partial_{\mu}\alpha(t,x)$ and $\psi'(t,x)=\exp(-\rmi \alpha(t,x))\psi(t,x)$,
where $\alpha(t,x)$ is an arbitrary spacetime dependent continuous rotation angle (defined mod $2\pi$).
In addition to the infinitesimally generated ``small gauge transformations'' continuously connected to the identity transformation
with $\alpha(t,x)=0$, spatial compactification brings to the fore the topologically non trivial group of ``large gauge transformations''.
The distinction between these classes of transformations is made explicit by expressing
the arbitrary function $\alpha(t,x)$ through the decomposition $\alpha(t,x)=\alpha_{0}(t,x)+2\pi x\ell/L$ in terms of
a periodic function $\alpha_{0}(t,x)=\alpha_{0}(t,x+L)$ and an integer $\ell \in \mathbb{Z}$, the so-called (additive) winding number
of the ``large gauge transformation''. This group of integers is the fundamental or first homotopy group $\pi_{1}(S^1)$ which classifies
the mappings $S^{1}\rightarrow U(1)$. ``Small gauge transformations'' form the local gauge group, \emph{i.e.}, they are connected
to the identity. These transformations are generated by exponentiation of the parameter $\alpha(t,x)=\alpha_0(t,x)$ with $\ell=0$.
If the holonomy of the gauge transformation around the circle, namely $\ell\in \mathbb{Z}$, does not vanish, we are dealing
with a large gauge transformation. One of the purposes of this paper is to emphasize the topological difference between these two classes
of gauge transformations and especially the consequences of large gauge transformations. This is done by considering
the ``modular group'', namely the quotient of the full gauge group
by the local gauge group. For the present system the modular group is isomorphic to the additive group $\mathbb{Z}$ of the winding
number $\ell$. It will be shown that complete gauge invariance under all gauge transformations may conveniently be enforced by
requiring separately invariance under the local gauge group and the modular group.
 
One may take advantage of these considerations to distinguish the various sectors on which these gauge transformations act.
From the point of view of the spatial $S^1$ which is a compact manifold, let us apply a Hodge decomposition of the gauge field
of which the time component is a 0-form and the space component a 1-form. Hence,
\begin{eqnarray}
A_{0}(t,x)=a_{0}(t)+\partial_{1}\omega_{1}(t,x),\\
A_{1}(t,x)=a_{1}(t)+\partial_{1}\phi(t,x),\label{Hodge}
\end{eqnarray}
where the periodic functions $\omega_{1}(t,x)$ and $\phi(t,x)$ do not include a spatial zero-mode, \emph{i.e.}, these 1- and 0-form
fields do not include a space independent component, while $a_0(t)$ and $a_1(t)$ are the corresponding harmonic forms.
Similarly a Hodge decomposition also applies to the gauge parameter 0-form,
\begin{eqnarray*}
\alpha_{0}(t,x)=\beta_{0}(t)+\partial_{1}\beta_{1}(t,x),
\end{eqnarray*}
where once again the 1-form $\beta_{1}(t,x)$ does not include a (spatial) zero-mode.
In terms of this separation of variables, gauge transformations of winding number $\ell$ and parameter
$\alpha(t,x)=\alpha_0(t,x)+2\pi x\ell/L$ act as follows on the Hodge components of $A_0(t,x)$,
\begin{eqnarray*}
\cases{a_{0}'(t)=a_{0}(t)+\frac{1}{e}\partial_{0}\beta_{0}(t),\\
\omega_{1}(t,x)=\omega_{1}(t,x)+\frac{1}{e}\partial_{0}\beta_{1}(t,x),}
\end{eqnarray*}
while for $A_{1}(t,x)$,
\begin{eqnarray*}
\cases{
a_{1}'(t)=a_{1}(t)+\frac{2\pi \ell}{eL},\label{largegaugea}\\
\phi'(t,x)=\phi(t,x)+\frac{1}{e}\partial_{1}\beta_{1}(t,x).}
\end{eqnarray*}

A noticeable fact is that the modular transformation of winding number $\ell$ is found to act in the gauge sector
only as a shift in the zero-mode $a_1(t)$ which is itself invariant under any local gauge transformation.
Furthermore the Hodge decomposition in \eref{Hodge} allows one to ``dress'' the fermionic field with the longitudinal gauge field as follows
\begin{eqnarray}
\chi(t,x)=\exp(\rmi e\phi(t,x))\psi(t,x).
\end{eqnarray}
This redefinition of the Dirac spinor is reminiscent of Dirac's construction \cite{Dirac:1955uv} of a ``physical electron''
carrying its own ``photon cloud'' so that this composite object be gauge invariant.
The boundary condition for the dressed fermion is still given by the holonomy condition of parameter $\lambda$,
$\chi(t,x+L)=\exp(-2\rmi \pi \lambda)\,\chi(t,x)$.
However gauge transformations of the redefined spinor simplify as,
\begin{eqnarray}
\chi'(t,x)'=\exp(- \rmi\beta_{0}(t))\ \exp(-2\rmi\pi \ell\frac{x}{L})\ \chi(t,x),\label{largegaugechi}
\end{eqnarray}
showing that a local gauge transformation induces only a time dependent but space independent
phase change $\exp(- \rmi\beta_{0}(t))$ of the ``composite'' fermionic field.
A space dependent gauge transformation of $\chi(t,x)$ is associated now to the modular group only, whose
topologically non trivial action multiplies $\chi(t,x)$ by $\exp(-2\rmi\pi \ell x/L)$.
In other words modular transformations, which account for the topological features of the compactified theory and its gauge symmetries,
act only on the following degrees of freedom,
\begin{eqnarray*}
\chi'(t,x)=\exp(-2\rmi\pi \ell\frac{x}{L})\ \chi(t,x),\qquad
a_{1}'(t)=a_{1}(t)+\frac{2\pi \ell}{eL},\qquad
\ell\in\mathbb{Z}.
\end{eqnarray*}

These different field redefinitions making manifest a separation of the gauge degrees of freedom into local and topological ones,
imply the following expression for the action of the theory,
\begin{eqnarray*}
S=\int dt&\Big\{\frac{1}{2}L\dot{a}_{1}^{2}-ea_{0}\int_{S^1} dx\chi^{\dagger}\chi-ea_{1}\int_{S^1} dx\overline{\chi}\gamma^{1}\chi\\
&+\int_{S^1} dx\Big(\frac{1}{2}\rmi\chi^{\dagger}\partial_{0}\chi-\frac{1}{2}\rmi\partial_{0}\chi^{\dagger}\chi+\frac{1}{2}\rmi\overline{\chi}\gamma^1\partial_{1}\chi-\frac{1}{2}\rmi\partial_{1}\overline{\chi}\gamma^1\chi-\mu\overline{\chi}\chi\\
&-\frac{1}{2}(\partial_{0}\phi-\partial_{1}\omega_{1})\partial_{1}^{2}(\partial_{0}\phi-\partial_{1}\omega_{1})+e(\partial_{0}\phi-\partial_{1}\omega_{1})(\chi^{\dagger}\chi)'\Big)\Big\},
\end{eqnarray*}
where the notation $(\chi^\dagger \chi)'$ stands for the quantity shown in parenthesis but with its spatial zero-mode subtracted
(and where as usual a dot above a quantity stands for the time derivative of that quantity).

Given the existence of gauge symmetries, the identification of the Hamiltonian formulation of this system must
rely on the methods of constrained dynamics \cite{Govaerts:1991gd}. The momenta canonically conjugate to all degrees of freedom are
(here Grassmann odd derivatives for the spinor components are left-derivatives, while $L_0$ is the total quantity in curly brackets in the
above expression for the action),
\begin{eqnarray*}
p^{0}=\frac{\partial L_0}{\partial \dot{a}_0}=0,\\
\pi^1=\frac{\partial L_0}{\partial \dot{\omega}_1}=0,\\
p^{1}=\frac{\partial L_0}{\partial \dot{a}_i}=L\dot{a}_{1}, \\
\pi_{\phi}=\frac{\partial L_0}{\partial \dot{\phi}}=-\bigtriangleup (\partial_{0}\phi-\partial_{1}\omega_{1})+e(\chi^{\dagger}\chi)',\\
\xi_{1}=\frac{\partial L_0}{\partial \dot{\chi}}=-\frac{1}{2}\rmi\chi^\dagger,\\
\xi_{2}=\frac{\partial L_0}{\partial \dot{\chi}^\dagger}=-\frac{1}{2}\rmi\chi ,
\end{eqnarray*}
with $\xi^{\dagger}_{1}(t,x)=-\xi_{2}(t,x)$. For two of the degrees of freedom one may express their velocity in terms of
their conjugate momentum, namely $\dot{a}_{1}(t)=p^{1}(t)/L$ and
$\partial_{0}\phi(t,x)=\partial_{1}\omega_{1}(t,x)-\bigtriangleup^{-1}(\pi_{\phi}(t,x)-e(\chi^\dagger \chi)'(t,x))$.
Here the symbol $\bigtriangleup^{-1}$ denotes the Green function of the spatial Laplacian on the circle, $\Delta=\partial^2_1$, 
again not including the spatial zero-mode. Since $\pi_{\phi}$ does not include a zero-mode the action of $\bigtriangleup^{-1}$
in the previous expression for $\partial_0\phi$ is well defined. However since the Hessian of the Lagrange function for the other
degrees of freedom possesses null eigenvectors, there exist primary phase space constraints.
Clearly these primary constraints are $p^{0}(t)=0$, $\pi^1(t,x)=0$, $\xi_{1}(t,x)+\rmi\chi^\dagger(t,x)/2=0$ and
$\xi_{2}(t,x)+\rmi\chi(t,x)/2=0$. 

Since the canonical Hamiltonian is readily identified to be given as,
\begin{eqnarray*}
H_{0}&=\frac{1}{2L}(p^1)^2+ea_{0}\int_{S^1}dx  \chi^{\dagger}\chi +ea_{1}\int_{S^1}dx \overline{\chi}\gamma^{1}\chi  +\\
&+\int_{S^1}dx\Big\{-\frac{1}{2}\rmi\overline{\chi}\gamma^{1}\partial_{1}\chi+\frac{1}{2}\rmi\partial_{1}\overline{\chi}\gamma^{1}\chi +\mu\overline{\chi}\chi\\
&+\partial_{1}\omega_{1}\pi_{\phi}-\frac{1}{2}(\pi_{\phi}-e(\chi^\dagger\chi))'\bigtriangleup^{-1}(\pi_{\phi}-e(\chi^\dagger\chi))'\Big\}.
\end{eqnarray*}
a consistent time evolution of the primary constraints must consider as primary Hamiltonian the following total quantity
\begin{eqnarray}
H_{1}=H_{0}+\lambda_{0}p^{0}+\int_{S^1}dx\Big(\lambda_{1}\pi^{1}+(\xi_{1}+\frac{1}{2}\rmi\chi^\dagger)\tilde{\lambda}_{1}+
\tilde{\lambda}_{2}(\xi_{2}+\frac{1}{2}\rmi\chi)\Big),
\end{eqnarray}
where $(\lambda_{0}(t)$, $\lambda_{1}(t,x))$ and $(\tilde{\lambda}_{1}(t,x)$, $\tilde{\lambda}_{2}(t,x))$ are 
Grassmann even and Grassmann odd would-be Lagrange multipliers, respectively. Requiring a consistent time evolution of
the primary constraints generated through the (Grassmann graded) Poisson brackets by this primary Hamiltonian implies
the following further conditions,
\begin{eqnarray*}
\{p^0,H_1\}=-e\int_{S^1}dx\chi^\dagger\chi=0, \\
\{\pi^1,H_1\}=\partial_{1}\pi_{\phi}=0,\\
\{\xi_{1}+\frac{1}{2}\rmi\chi^\dagger,H_1\}=0,\\
\{\xi_{2}+\frac{1}{2}\rmi\chi,H_1\}=0.
\end{eqnarray*}
In actual fact, the last two conditions imply equations for the Grassmann odd multipliers $\tilde{\lambda}_{1}$ and $\tilde{\lambda}_{2}$
which are thereby uniquely determined. The other two conditions however, define secondary constraints, the first of which,
namely $e\int_{S^1}dx\chi^\dagger\chi=0$, is the zero-mode of the ordinary Gauss law. A consistent time evolution
of these new constraints requires to include them in a secondary Hamiltonian which is to generate time evolution,
\begin{eqnarray}
H_{2}=H_{1}+e\lambda_3\int_{S^1}dx\chi^\dagger\chi+\int_{S^1}dx\lambda^1_{3}\partial_{1}\pi_{\phi},
\end{eqnarray}
where $\lambda_3(t)$ and $\lambda^1_{3}(t,x)$ are would-be Lagrange multipliers enforcing the secondary constraints.
It is readily checked that no further constraints are then generated from $H_2$. A consistent time evolution of
physical states is ensured.

According to Dirac's classification the set of constraints decomposes into first and second class constraints.
In the case under study, $p^0=0$ and $e\int_{S^1}dx\chi^\dagger\chi=0$ are first class while
$\xi_{1}+\frac{1}{2}\rmi\chi^\dagger=0$ and $\xi_{2}+\frac{1}{2}\rmi\chi=0$ are second class constraints.
First class constraints always generate gauge symmetries. Second class constraints on the other hand, indicate
that some degrees of freedom are unnecessary and may be reduced through the introduction of the associated
Dirac brackets. In the present case Dirac brackets act in the fermionic sector only, and are given as,
\begin{eqnarray}
\big\{\chi_{\alpha}(t,x),\chi^{\dagger}_{\beta}(t,y)\big\}_{D}=-\rmi\delta_{\alpha,\beta}\delta_{S^{1}}(x-y)
\exp\left(-2 \rmi \pi\frac{(x-y)}{L}\lambda\right),
\end{eqnarray}
where $\lambda$ is the fermionic holonomy while $\delta_{S^1}(x-y)$ stands for the Dirac $\delta$-function defined over
the spatial circle $S^1$, and $\alpha,\beta=1,2$ are spinor indices.

The first-order action associated with the Hamiltonian formulation is thus defined by the first-order Lagrange functional
\begin{eqnarray*}
L=&\dot{a}_{0}p^0+\dot{a}_{1}p^{1}-\lambda_{0}p^0
-ea_{1}\int_{S^1}dx\overline{\chi}\gamma^{1}\chi-e(a_0+\lambda_3)\int_{S^1}dx\chi^\dagger\chi\\
&-\frac{p_{1}^2}{2L}+\int_{S^1}dx\Big\{\partial_0\omega_{1}\pi^1+\partial_{0}\phi\pi_{\phi}-\partial_{1}\omega_{1}\pi_{\phi}-\lambda_{3}^{1}\partial_{1}\pi_{\phi}-\lambda_{1}\pi^{1}\\
&+\frac{1}{2}\rmi\chi^\dagger\partial_{0}\chi-\frac{1}{2}\rmi\partial_{0}\chi^\dagger\chi
+\frac{1}{2}\rmi\overline{\chi}\gamma^{1}\partial_{1}\chi
-\frac{1}{2}\rmi\partial_{1}\overline{\chi}\gamma^{1}\chi-\mu\overline{\chi}\chi\\
&+\frac{1}{2}(\pi_{\phi}-e(\chi^\dagger\chi)')\bigtriangleup^{-1}(\pi_{\phi}-e(\chi^\dagger\chi))'\Big\}.
\end{eqnarray*}
However some of the first class constraints, namely $p^0=0$ and $\pi^1=0$, appear because some of the degrees of freedom
are in actual fact already Lagrange multipliers for some of the other first class constraints, namely in the present
case $A_0(t,x)=a_0(t)+\partial_1\omega_1(t,x)$ is the Lagrange multiplier for Gauss' law which is the first class constraint
generating small gauge transformations of parameter $\alpha_0(t,x)$. In such a situation one may use the freedom in choosing the Lagrange
multipliers for such superfluous first class constraints without affecting the actual gauge invariances of the system,
and thereby determine a more ``fundamental" or basic Hamiltonian formulation \cite{Govaerts:1991gd}.
First let us make the choice $\lambda_0(t)=\dot{a}_{0}(t)$ and then replace $a_0(t)+\lambda_3(t)$ by $a_0(t)$.
Consequently the sector $(a_{0},p^{0})$ decouples altogether from the dynamics, with the new variable $a_0(t)$ being the Lagrange multiplier
for the first class constraint $e\int_{S^1}dx\chi^\dagger\chi=0$.
Likewise the choice $\lambda_{1}(t,x)=\partial_0\omega_{1}(t,x)$ and then applying the redefinition
$-\lambda^{1}_{3}(t,x)+\omega_{1}(t,x)\rightarrow \lambda^1(t,x)$ shows that the sector $(\omega_{1},\pi^{1})$ decouples as well,
with the new quantity $\lambda^1(t,x)$ being the Lagrange multiplier for the first class constraint $\partial_1\pi_\phi=0$.
Given these redefinitions the Hamiltonian formulation is specified by the first order Lagrangian
\begin{eqnarray*}
L=&\dot{a}_{1}p^{1}-\frac{1}{2L}(p^{1})^{2}-ea_0\int_{S^1}dx\chi^\dagger\chi-ea_{1}\int_{S^1}dx\overline{\chi}\gamma^{1}\chi\nonumber\\
&+\int_{S^1}dx\Big\{\partial_{0}\phi\pi_{\phi}+\frac{1}{2}\rmi\chi^\dagger\partial_{0}\chi-\frac{1}{2}\rmi\partial_{0}\chi^\dagger\chi
+\frac{1}{2}\rmi\overline{\chi}\gamma^{1}\partial_{1}\chi-\frac{1}{2}\rmi\partial_{1}\overline{\chi}\gamma^{1}\chi\nonumber\\
&+\lambda^1\partial_{1}\pi_{\phi}-\mu\overline{\chi}\chi+\frac{1}{2}(\pi_{\phi}-e(\chi^\dagger\chi))'\bigtriangleup^{-1}(\pi_{\phi}-e(\chi^\dagger\chi))'\Big\}.
\end{eqnarray*}
However, since the sector $(\phi,\pi_\phi)$ contributes only linearly and quadratically to this action,
it may easily be reduced as well through its equations of motion, which read,
\begin{eqnarray*}
\cases{
\partial_{0}\phi=-\bigtriangleup^{-1}(\pi_{\phi}-e(\chi^\dagger\chi)')+\partial_{1}\lambda^{1},\\
\partial_{0}\pi_{\phi}=0,}
\end{eqnarray*}
with the constraint $\partial_1\pi_\phi=0$, where $\pi_{\phi}$ does not include a zero-mode.
Hence one has $\pi_\phi(t,x)=0$ while the pure gauge degree of freedom $\phi(t,x)$ is determined
from $\partial_0\phi=e\Delta^{-1}(\chi^\dagger\chi)'+\partial_1\lambda^1$.

Upon this final reduction, the Hamiltonian formulation of the system consists of the
phase space variables $(a_1(t),p^1(t);\chi(t,x),\chi^\dagger(t,x))$ with the Poisson-Dirac brackets
\begin{eqnarray*}
\left\{a_1(t),p^1(t)\right\}=1,\\
\big\{\chi_{\alpha}(t,x),\chi^{\dagger}_{\beta}(t,y)\big\}_{D}=-\rmi\delta_{\alpha,\beta}\delta_{S^{1}}(x-y)\exp\left(-2 \rmi \pi(x-y)\lambda\right),
\end{eqnarray*}
subjected to the single first class constraint $e\int_{S^1}dx\chi^\dagger\chi=0$ of which the Lagrange multiplier is $a_0(t)$,
and a dynamics deriving from the Hamiltonian first-order action
\begin{eqnarray*}
S=&\int dt\Big\{\dot{a}_1p^1+\int_{S^1}dx\Big(\frac{1}{2}\rmi\chi^\dagger\partial_{0}\chi-\frac{1}{2}\rmi\partial_{0}\chi^\dagger\chi\Big)\\
&\,-\,H\,-\,ea_0\int_{S^1}dx \chi^\dagger\chi\Big\},
\end{eqnarray*}
where the first class Hamiltonian $H$ is given by,
\begin{eqnarray}
H=&\frac{(p^{1})^{2}}{2L} \nonumber \\
&+\int_{S^{1}}dx\Big\{\overline{\chi}\gamma^{1}(-\rmi\partial_{1}+ea_{1})\chi+\mu\overline{\chi}\chi
-\frac{1}{2}e^{2}(\chi^{\dagger}\chi)'\bigtriangleup^{-1}(\chi^{\dagger}\chi)'\Big\}.
\label{Hbasic}
\end{eqnarray}
Note how the very last four-fermion contribution to $H$ stands for the instantaneous Coulomb interaction, even though no
gauge fixing procedure has been enforced, but rather a parametrization of the degrees of freedom which factorizes the
physical from the gauge dependent degrees of freedom.
The remaining gauge invariances of the system in the present formulation consist of the space independent small gauge
transformations with parameter $\alpha_0(t,x)=\beta_0(t)$ which are generated by the single remaining first class constraint,
$e\int_{S^1}dx \chi^\dagger\chi=0$, as well as the modular transformations of winding numbers $\ell\in\mathbb{Z}$, acting
as follows on the phase space variables,
\begin{eqnarray}
a'_1(t)=a_1(t)+\frac{2\pi\ell}{eL},\nonumber\\
{p^1}'(t)=p^1(t),\label{eq:survgauge}\\
\chi'(t,x)=\exp(-i\beta_0(t))\ \exp(-2i\pi\ell\frac{x}{L})\ \chi(t,x).\nonumber
\end{eqnarray}
In particular the first class constraint, merely the space integrated Gauss law, requires physical states to carry
a vanishing net electric charge. In addition however, physical states need also to be modular invariant, a restriction which is
intrinsically of a purely topological character involving the gauge harmonic form $a_1(t)$ as well as the winding numbers of
the gauge symmetry group.

\section{Canonical Quantisation}

Canonical quantization of the system in the Schr\"odinger picture (at $t=0$) proceeds from its basic Hamiltonian formulation
of the previous Section. It is necessary to consider a mode expansion of the dressed spinor $\chi(t=0,x)$, which is taken in the form,
\begin{eqnarray}
\chi(x)=\sqrt{\frac{\hbar}{L}}\sum_{m\in\mathbb{Z}}\left(\begin{array}{cc}
d^{\dagger}_{-m}\\
b_{m}
\end{array}\right)
\exp(2\rmi\pi\frac{x}{L}(m-\lambda)),
\end{eqnarray}
with the anticommutation relations
$\{d_{-m},d^{\dagger}_{-n}\}=\delta_{m,n}=\{b_{m},b^{\dagger}_{n}\}$. Note that the mode indices $m,n\in\mathbb{Z}$ also label
the momentum eigenvalues $2\pi m/L$ of the fermion total momentum operator. For example $b_m$ and $d^\dagger_{-m}$ both carry
momentum $(-2\pi m/L)$. A particle and anti-particle interpretation of the sectors $(b_m,b^\dagger_m)$ and $(d_m,d^\dagger_m)$,
respectively, is warranted by considering the mode expansion of the total electric charge, $Q=\int_{S^1}dx\chi^\dagger(x)\chi(x)$
(the specific definition and expression of this composite operator is provided below). This choice of mode expansion translates
also into the following anticommutation relations for the spinor field,
\begin{eqnarray*}
\{\chi_{\alpha}(x),\chi^\dagger_{\beta}(y)\}=\delta_{\alpha,\beta}\frac{\hbar}{L}\sum_{m}\exp(2\rmi\pi\frac{x-y}{L}(m-\lambda)),
\end{eqnarray*}
which are in direct correspondence with their classical Dirac bracket counterparts. Similarly, the zero-mode of the gauge sector, $(a_{1},p^{1})$,
is quantized by the Heisenberg algebra, $\big[\hat{a}_{1},\hat{p}^{1}\big]=\rmi\hbar$, $\hat{a}_1$ and $\hat{p}^1$ needing
to be self-adjoint operators as well.

In terms of the above mode expansion the fermionic bilinear contribution to the first class Hamiltonian \eref{Hbasic},
namely $H=(p^1)^2/(2L)+H_0+H_C$, takes the form
\begin{eqnarray*}
H_0&=&\int_{S^{1}}dx\overline{\chi}\gamma^{1}(-\rmi\partial_{1}+e\hat{a}_{1})\chi\\
&=&\sum_{m}\Big((2\pi\frac{m-\lambda}{L}+e\hat{a}_{1})(b^{\dagger}_{m}b_{m}-d_{-m}d^{\dagger}_{-m})+\mu(d_{-m}b_{m}+b^{\dagger}_{m}d^{\dagger}_{-m})\Big),
\end{eqnarray*}
while the instantaneous Coulomb interaction energy becomes,
\begin{eqnarray*}
H_{C}=\kappa\sum_{\ell\neq 0}\frac{1}{\ell^{2}}\Big(\sum_{m,n}(d_{-n}d^{\dagger}_{-m}+b^{\dagger}_{n}b_{m})
\delta_{m,n+\ell}\Big)\Big(\sum_{p,q}(d_{-q}d^{\dagger}_{-p}+b^{\dagger}_{q}b_{p})\delta_{p,q-\ell}\Big),
\label{eq:HC}
\end{eqnarray*}
with $\kappa=e^{2}L/(2(2\pi)^{2})$.
To establish the last expression the following representation of the Green function of the spatial Laplacian is used, 
\begin{eqnarray*}
(\bigtriangleup^{-1}g)(x)=\frac{-1}{L}\int_{S^{1}}dy\sum_{\ell\neq 0}\frac{\exp(2\rmi\pi(x-y)\frac{\ell}{L})}{(\frac{2\pi \ell}{L})^2}g(y).
\end{eqnarray*}
Note that a specific ordering prescription for these composite operators $H_0$ and $H_C$ is implicit at this stage.
An explicit ordering prescription and complete definition of composite operators is to be given hereafter.

A consistent quantization should also implement the action of all remaining gauge transformations, in correspondence with
the classical transformations \eref{eq:survgauge}, through the adjoint action of specific quantum operators.
The action of the modular transformation of winding number $\ell$ is 
\begin{eqnarray}
  \hat{U}(\ell)\,\hat{a}_1\,\hat{U}^{\dag}(\ell)&=\hat{a}_1+\frac{2\pi}{eL}\ell,&\qquad
\hat{U}(\ell)\,\hat{p}^1\,\hat{U}^\dagger(\ell)=\hat{p}^1,\nonumber\\
  \hat{U}(\ell)\,b_{m}\,\hat{U}^{\dag}(\ell)&= b_{m+\ell},&\qquad
  \hat{U}(\ell)\,d_{-m}\,\hat{U}^{\dag}(\ell)= d_{-m-\ell},
\label{largegauge}
\end{eqnarray}
with the corresponding quantum modular operator of winding number $\ell\in \mathbb{Z}$ given as,
\begin{eqnarray}
\hat{U}(\ell)=\exp\Big\{ 2\rmi \pi
\ell\Big(\frac{1}{e}\frac{\hat{p}_{1}}{L}-\frac{\theta_0}{2\pi}
+\frac{1}{L}\int_{S_{1}}dx:x\chi^{\dag}(x)\chi(x):\Big)\Big\}.
\end{eqnarray}
The actual meaning of the ordering prescription, ``$: \ :$'', is specified below.
The arbitrary new constant parameter $\theta_0$, which is defined mod $2\pi$, arises as follows.
The quantum unitary operators, $\hat{U}(\ell)$, realising modular transformations involve {\it a priori}
an arbitrary phase factor that may be winding number dependent. However since the modular group is additive
in the winding number, the choice of phase should be consistent with the group composition law,
$\hat{U}(\ell_1)\,\hat{U}(\ell_2)=\hat{U}(\ell_1+\ell_2)$. The general solution to this requirement implies
that the phase factor be linear in the winding number, hence the $\theta_0$ parameter as the arbitrary linear factor in $\ell$.
In actual fact, $\theta_0$ may be viewed as defining a purely quantum mechanical degree of freedom \cite{Govaerts:1999ep,FPayen},
and is the analogue for the present model of the $\theta$ vacuum angle in QCD.

Similarly small gauge transformations act as follows
\begin{eqnarray*}
b_{m}&\rightarrow \exp(-i \beta_0)b_m,\qquad
d_{-m} &\rightarrow \exp(i \beta_0)d_{-m},\nonumber
\end{eqnarray*}
while the corresponding quantum generator, namely the total electric charge $Q$ which is the first class constraint for these
local symmetries and of which the exponential, when multiplied by a factor proportional to $\beta_0$, determines the unitary operator
of which the adjoint action induces these finite transformations, is defined hereafter.

What is most remarkable indeed about these modular transformations is that in the fermionic sector they map spinor modes of a given
electric charge and of all possible momentum values into one another. In other words, modular symmetries, which are characteristic
of the topological properties of a gauge invariant system, induce transformations connecting the infrared and the ultraviolet,
namely the large and the small distance properties of a gauge invariant dynamics. This observation remains totally relevant in the
context of non-abelian Yang-Mills theories as well, coupled to charge matter fields. Physical consequences of such modular symmetries
are presumably far reaching, and deserve to be fully explored especially since they are intrinsically of a topological hence
non-perturbative character.

Obviously composite quantum operators need to be carefully defined in order to preserve the modular gauge symmetry in a manifest way
(see \eref{largegauge}; that a regularization prescription also preserves in a manifest way gauge invariance under local small transformations
is readily checked). Let us first consider the bilinear fermion contributions to the first class Hamiltonian $H$, which need to
be properly defined to ensure both finite matrix elements and a ground state of finite energy, given that $b_m$ and $d_m$ are
taken to be annihilators of a fermionic Fock vacuum, with $b^\dagger_m$ and $d^\dagger_m$ acting as creators. Making the choice\footnote{Other
regularization choices have been considered, and shown to lead to the same final conclusions.}
of a gaussian regularization with energy cut-off $\Lambda$, the bilinear fermion contributions to the first class Hamiltonian become,
\clearpage
\begin{eqnarray*}
\sum_{m}&\Big\{\left(\frac{2\pi}{L}(m-\lambda)+e\hat{a}_{1}\right)
\left(b_{m}^{\dagger}b_{m}-d_{-m}d_{-m}^{\dagger}\right)+\mu \left(d_{-m}b_{m}+b^{\dagger}_{m}d^{\dagger}_{-m}\right)\Big\}\\
&\times \exp\left(-\frac{1}{\Lambda^2} \left(\frac{2 \pi}{L}(m-\lambda)+e\hat{a}_{1}\right)^2\right).
\end{eqnarray*}
This choice of regularization prescription ensures that this bilinear operator has finite matrix elements while it remains manifestly invariant
under all modular gauge transformations~\eref{largegauge}. A further subtraction to be discussed hereafter, still needs to be applied to
this expression, in order that eventually the regulator may be removed while leaving a well defined composite operator $H_0$.
Let us note that the mass term couples left- and right-moving modes. This fact will make possible to smoothly redefine what will
be the creators and annihilators of left- and right-moving particles.

In order to diagonalize this regularized operator, let us consider the sector of modes $(b_m,b^\dagger_m)\equiv(b,b^\dagger)$ and
$(d_{-m},d^\dagger_{-m})\equiv(d,d^\dagger)$ for any given $m\in\mathbb{Z}$. For definiteness the corresponding fermionic Fock space
is spanned by the Fock vacuum $|0,0\rangle$ and the states $|1,0\rangle=b^{\dagger}|0,0\rangle$, $|0,1\rangle=d^{\dagger}|0,0\rangle$ and
$|1,1\rangle=d^{\dagger}b^{\dagger}|0,0\rangle$. The contribution of that sector to the above bilinear operator is thus
of the following form,
\begin{eqnarray*}
h=\beta (b^{\dagger}b-dd^{\dagger})+\alpha (b^{\dagger}d^{\dagger}+db),
\end{eqnarray*}
with $\beta=(\frac{2\pi}{L}(m-\lambda)+e\hat{a}_{1})\exp\{-(\frac{2 \pi}{L}(m-\lambda)+e\hat{a}_{1})^2/\Lambda^2\}$ and
$\alpha=\mu \exp\{-(\frac{2 \pi}{L}(m-\lambda)+e\hat{a}_{1})^2/\Lambda^2\}$.
This operator $h$ has 4 orthonormalized eigenstates listed in Table 1, in which
$\psi_{\mp}=-(\beta \pm\sqrt{\beta^2+\alpha^2})/\alpha$ so that $\psi_{+}\psi_{-}=-1$. 

\begin{table}[h!]
\caption{\label{eigenstates}Eigenstates and eigenvalues of $h$.}
\begin{tabular*}{\textwidth}{@{}l*{15}{@{\extracolsep{0pt plus
12pt}}l}}
\br
State&Eigenvalue\\
\mr
$|\psi_{+}\rangle=\frac{|0,0\rangle+\psi_+|1,1\rangle}{\sqrt{1+\psi_+^2}}$  & $\sqrt{\alpha^2+\beta^2}$ \\ 
$|1,0\rangle$ \mbox{ and } $|0,1\rangle$ &  $0$\\ 
$|\psi_{-}\rangle=\frac{|0,0\rangle+\psi_-|1,1\rangle}{\sqrt{1+\psi_-^2}}$  & $-\sqrt{\alpha^2+\beta^2} $\\
\br
\end{tabular*}
\end{table}

In any given $m$ sector, the state $|\psi_{-}\rangle$ is thus the minimal energy eigenstate.
One may consider two pairs of fermionic creators and annihilators defined by
\begin{eqnarray*} 
B^\dagger_{\pm}=\frac{b^\dagger+\psi_{\pm}d}{\sqrt{1+\psi_{\pm}^2}},\qquad
D_{\pm}=\frac{d-\psi_{\pm}b^{\dagger}}{\sqrt{1+\psi_{\pm}^2}},
\end{eqnarray*}
whether for the index ``$+$'' or the index ``$-$''.
These $B$ and $D$ operators and their adjoints obey two separate fermionic Fock algebras
whether for the index ``$+$'' or the index ``$-$'', namely $\{B^\dagger_{\pm,m},B_{\pm,n}\}=\delta_{m,n}$ and
$\{D^\dagger_{\pm,-m},D_{\pm,-n}\}=\delta_{m,n}$.
The operators $B_+$ and $D_+$ (resp., $B_-$ and $D_-$) annihilate the state $|\psi_{+}\rangle$ (resp., $|\psi_{-}\rangle$).
Given these definitions, $h$ acquires two separate though equivalent expressions,
 \begin{eqnarray*}
h=-\sqrt{\alpha^2+\beta^2}\big(B^{\dagger}_{+}B_{+}-D_{+}D^{\dagger}_{+}\big)=
\sqrt{\alpha^2+\beta^2}\big(B^{\dagger}_{-}B_{-}-D_{-}D^{\dagger}_{-}\big).
\end{eqnarray*}
Among these two possibilities, in the sequel let us choose to work with the operators defined with the ``$-$'' index, of which
$B_-$ and $D_-$ thus annihilate the ground state in the fermionic sector $m$, $|\psi_{-}\rangle$,
\begin{eqnarray*}
B_{-}|\psi_{-}\rangle=0=D_{-}|\psi_{-}\rangle.
\end{eqnarray*}
Henceforth the index ``$-$'' will thus be suppressed, with $(B_m,B^\dagger_m)$ and
$(D_{-m},D^\dagger_{-m})$ acting truly as annihilators and creators of fermionic Fock algebras of which the
Fock vacuum is the state $|\psi_-\rangle$. Note however that all these quantities involve also
the gauge zero-mode operator $\hat{a}_1$.

We may now rewrite all the quantities of interest in terms of the original variables,
\begin{eqnarray*}
{\ }\hspace{-10pt}
\sqrt{\alpha^2+\beta^2}=\big[(\frac{2 \pi}{L}(m-\lambda)+e\hat{a}_{1})^2+\mu^2\big]^{1/2}
\exp\big[-\frac{1}{\Lambda^2} (\frac{2 \pi}{L}(m-\lambda)+e\hat{a}_{1})^2\big],
\end{eqnarray*}
and
\begin{eqnarray*}
\psi_{-}=-\frac{\frac{2 \pi}{L}(m-\lambda)+e\hat{a}_{1}}{\mu}+\frac{1}{\mu}\sqrt{(\frac{2 \pi}{L}(m-\lambda)+e\hat{a}_{1})^2+\mu^2}.
\end{eqnarray*}
It is convenient to introduce a rotation angle, so that $\cos \phi_{-}=1/\sqrt{1+\psi_{-}^2}$ and $\sin \phi_{-}=\psi_{-}/\sqrt{1+\psi_{-}^2}$.

Consider now the limit where $\mu$ tends to zero. First, if $\frac{2\pi}{L}(m-\lambda)+e\hat{a}_{1} \neq 0$ the limit $\mu\rightarrow 0$ implies
\begin{eqnarray*}
\lim_{\mu\to 0}
\left(\begin{array}{c}
b^{\dagger}_{m}\\ 
d_{-m}
\end{array}\right) =
\left(\begin{array}{cc}
 \cos \phi_{-} & \sin \phi_{-}\\ 
-\sin \phi_{-} & \cos \phi_{-}
\end{array} \right)
\left(\begin{array}{cc}
B^{\dagger}_{m}\\ 
D _{-m}
\end{array}\right) ,
\end{eqnarray*}
with the following specific values,
\begin{eqnarray*}
\cos \phi_{-} =\cases{
1 \qquad \mbox{ if }   \frac{2 \pi}{L}(m-\lambda)+e\hat{a}_{1}>0;\\
0 \qquad \mbox{ if }    \frac{2 \pi}{L}(m-\lambda)+e\hat{a}_{1}<0;}
\end{eqnarray*}
\begin{eqnarray*}
\sin \phi_{-} =\cases{
0 \qquad \mbox{ if }    \frac{2 \pi}{L}(m-\lambda)+e\hat{a}_{1}>0;\\
1 \qquad \mbox{ if }     \frac{2 \pi}{L}(m-\lambda)+e\hat{a}_{1}<0.}
\end{eqnarray*}
If however $\frac{2 \pi}{L}(m-\lambda)+e\hat{a}_{1}=0$ the ``mixing angle'' is of $\pi/4$ radians in the massless limit,  
\begin{eqnarray*}
\lim_{\mu\to 0}
\left(\begin{array}{cc}
b^{\dagger}_{m}\\ 
d_{-m}
\end{array}\right) =
\left(\begin{array}{cc}
 \frac{1}{\sqrt{2}} & \frac{1}{\sqrt{2}}\\ 
\frac{-1}{\sqrt{2}} & \frac{1}{\sqrt{2}}
\end{array} \right)
\left(\begin{array}{c}
B^{\dagger}_{m}\\ 
D _{-m}
\end{array} \right).
\end{eqnarray*}
It is rather obvious that one may readily express all these results in terms of the Heaviside step function, $\Theta(x)$,
with the value $\Theta(0)=1/2$ as it turns out to be convenient for our purposes. However care needs to be exercised,
as the sequel will illustrate. It is also useful to note that
\begin{eqnarray*}
\Theta\big(\frac{2 \pi}{L}(m-\lambda)+e\hat{a}_{1}\big)=\Theta(m+\hat{a}),
\end{eqnarray*}
with the notation $\hat{a}=e\hat{a}_{1}L/(2\pi)-\lambda$.
Under a large gauge transformation of winding number $\ell$, $\hat{a}$ transforms as $\hat{a}\rightarrow \hat{a}+\ell$.
Finally we are in the position to make the following crucial identifications,
\begin{eqnarray*}
&\lim_{\mu \to 0} \cos \phi_{-} =\sqrt{\Theta(m+\hat{a})},\qquad
\lim_{\mu \to 0} \sin \phi_{-} =\sqrt{\Theta(-m-\hat{a})}.
\end{eqnarray*}

The above transformations ``\`a la Bogoliubov'' redefine creators and annihilators for Fock algebras through linear transformations. 
By construction this definition behaves ``covariantly'' under modular transformations, and may be written in a compact way as,
\begin{eqnarray}
&b^{\dagger}_{m}=B^{\dagger}_{m}\sqrt{\Theta(m+\hat{a})}+D_{-m}\sqrt{\Theta(-m-\hat{a})},\label{2un}\\ 
&d_{-m}=D_{-m}\sqrt{\Theta(m+\hat{a})}-B^{\dagger}_{m}\sqrt{\Theta(-m-\hat{a})}.\label{2deux}
\end{eqnarray}
while $d^{\dagger}_{-m}$ and $b_{m}$ are the adjoint operators of the previous expressions.
It is recalled also that $B^{(\dagger)}_m$ and $D^{(\dagger)}_{-m}$ involve an implicit dependence on $\hat{a}$.
The dependence on $\hat{a}$ of these definitions, with a spectral flow in the eigenvalues of that operator,
may be interpreted as a dynamical ``Fermi surface" in one dimension.

With the help of this definition, the electric charge operator reads,
\begin{eqnarray*}
Q=\sum_{m=-\infty}^{+\infty}\big(b^{\dagger}_{m}b_{m}+d_{-m}d^{\dagger}_{-m}\big)=\sum_{m=-\infty}^{+\infty}\big(B^{\dagger}_{m}B_{m}+D_{-m}D^{\dagger}_{-m}\big).
\end{eqnarray*}
An ordered expression of the gauge invariant regularized charge operator is, with $\tilde{\alpha}=2\pi/(L\Lambda^2)$,
  \begin{eqnarray*}
 :Q:\ \stackrel{\tilde{\alpha}\rightarrow 0}{=}\
\sum_{m=-\infty}^{+\infty}(B^{\dagger}_{m}B_{m}-D^{\dagger}_{-m}D_{-m}+1)\exp[-\tilde{\alpha}(m+\hat{a})^2],
  \end{eqnarray*}
where the divergent contribution independent of $\hat{a}$ may be subtracted while no further finite contribution in $\hat{a}$ arises.
The reader will find a detailed discussion of the technical result concerning the subtraction of infinities in \eref{divbil} of the Appendix.
In order to prove that no additional term depending on $\hat{a}$ is generated by the normal ordering procedure the Poisson resummation formula
is used, leading to
\begin{eqnarray}
{\ }\hspace{-25pt}
\sum_{m=-\infty}^{+\infty}\Theta(m+\hat{a})\exp(-\tilde{\alpha}(m+\hat{a})^2)\ \stackrel{\tilde{\alpha}\rightarrow 0}{=}
\ \frac{1}{2}\sqrt{\frac{\pi}{\tilde{\alpha}}}+ \sum_{n=-\infty,n\ne 0}^{+\infty}\frac{\exp(2\rmi\pi n\hat{a})}{2\pi\rmi n},\label{somme1}\\
{\ }\hspace{-25pt}
\sum_{m=-\infty}^{+\infty}\Theta(-m-\hat{a})\exp(-\tilde{\alpha}(m+\hat{a})^2)\ \stackrel{\tilde{\alpha}\rightarrow 0}{=}
\ \frac{1}{2}\sqrt{\frac{\pi}{\tilde{\alpha}}}+ \sum_{n=-\infty,n\ne 0}^{+\infty}\frac{\exp(2\rmi\pi n\hat{a})}{-2\pi\rmi n}.\label{somme2}
 \end{eqnarray} 
The subtraction consists in removing the contribution in $\frac{1}{2}\sqrt{\pi/\tilde{\alpha}}$ while no other infinite term remains. 
Eventually the normal ordered expression is given by
\begin{equation}
:\hat{Q}:_{\hat{a}}=\sum_{m=-\infty}^{+\infty}(B^{\dagger}_{m}B_{m}-D^{\dagger}_{-m}D_{-m}),
\end{equation} 
which is the definition of the quantum U(1) charge operator. The normal ordering prescription, $ : \ :_{\hat{a}}$,
depends on $\hat{a}$ in such a manner that this operation respects all gauge symmetries including modular transformations.
The regulator has safely been removed. As expected this operator is the generator of the $U(1)$ local gauge transformation,
\begin{eqnarray*}
B^{\dagger}_m &\rightarrow \exp(\rmi \beta)B^{\dagger}_m,\qquad
D^{\dagger}_{-m} &\rightarrow \exp(-\rmi \beta)D^{\dagger}_{-m}.\nonumber
\end{eqnarray*}

We may follow a similar analysis towards a quantum definition of the fermion bilinear contributions to the
first class Hamiltonian in the massless limit\footnote{A discussion of the modular invariant definition of this specific operator, in the context of the Schwinger model in the limit $e\to \infty$, is available in \cite{Itoi:1991ch}.},
 \begin{eqnarray*}
H_{bil}=\frac{2\pi}{L}\sum_{m=-\infty}^{+\infty}(m+\hat{a})(b^{\dagger}_{m}b_{m}-d_{-m}d^{\dagger}_{-m}).
 \end{eqnarray*}
With the help of the relations \eref{2un} and \eref{2deux}, the regularized normal ordered expression is,
 \begin{eqnarray*}
:H_{bil}:_{\hat{a}}=\frac{2\pi}{L}\sum_{m=-\infty}^{+\infty}|m+\hat{a}|(B^{\dagger}_{m}B_{m}+D^{\dagger}_{-m}D_{-m}-1)
\exp[-\tilde{\alpha}(m+\hat{a})^2].
 \end{eqnarray*}
Given the normal ordering contribution, the spectrum of $:H_{bil}:_{\hat{a}}$ includes an infinite contribution when the regulator is
removed. However we are not allowed to simply subtract this (regularized) contribution since it also
involves a dependence on $\hat{a}$, which is brought about by the choice of a modular invariant regularization.
The finite $\hat{a}$ dependent part may be computed after careful subtraction of the divergent contribution for $\hat{a}=0$.
Once again the Poisson resummation formula is used to isolate and extract the $\hat{a}$ dependent finite contribution.
Given \eref{divbis} in the Appendix, one finds
\begin{eqnarray*}
-\sum_{m=-\infty}^{+\infty}|m+\hat{a}|\exp[-\tilde{\alpha}(m+\hat{a})^2]\ \stackrel{\tilde{\alpha}\rightarrow 0}{=}
-\Big[\frac{2}{2\tilde{\alpha}}-2\sum_{n=-\infty,n\ne 0}^{+\infty}\frac{\exp(2\rmi\pi n \hat{a})}{(2\pi n)^2}\Big].
\end{eqnarray*}
The only divergence in $2/(2\tilde{\alpha})$ and which is independent of $\hat{a}$, is subtracted before removing the gaussian regulator. 
Thus finally the definition of this gauge invariant operator is,
\begin{eqnarray}
:\hat{H}_{bil}:_{\hat{a}}&=\frac{2\pi}{L}(\hat{a}-\lfloor \hat{a}\rfloor -\frac{1}{2})^2-\frac{\pi}{6L}\nonumber\\
&+\frac{2\pi}{L}\sum_{m=-\infty}^{+\infty}|m+\hat{a}|(B^{\dagger}_{m}B_{m}+D^{\dagger}_{-m}D_{-m}),
 \end{eqnarray}
where it is noted that the additional $\hat{a}$ dependent part is the Fourier series of a periodic potential given by
 \begin{eqnarray}
\sum_{n=-\infty,n\ne 0}^{+\infty}\frac{\exp(2\rmi\pi n a)}{(2\pi n)^2}=\frac{1}{2}(a-\lfloor a\rfloor -\frac{1}{2})^2-\frac{1}{24},
\end{eqnarray}
and where $\lfloor a \rfloor$ denotes the ``integer part'' of $a$, {\it i.e.}, the largest integer less or equal to $a$.
The quantum operator is bounded from below and is manifestly invariant under small as well as modular gauge transformations.
It is also relevant to address a well-known feature of the massless classical theory, namely its invariance under global chiral transformations, 
 \begin{eqnarray*}
b^{\dagger}_m &\rightarrow \exp(i \beta)b^{\dagger}_m,\qquad
d^{\dagger}_{-m} &\rightarrow \exp(i \beta)d^{\dagger}_{-m},\nonumber
\end{eqnarray*}
a symmetry which implies that the dynamics does not couple the left- and right-moving modes.
The corresponding classical conserved charge is the axial charge, which in the quantized theory takes the form,
\begin{eqnarray*}
{\ }\hspace{-20pt}
Q_{5}=\sum_{m=-\infty}^{+\infty}(b^{\dagger}_{m}b_{m}-d_{-m}d^{\dagger}_{-m})\\
{\ }\hspace{-20pt}
=\sum_{m=-\infty}^{+\infty}\Big\{sign(m+\hat{a})(B^{\dagger}_{m}B_{m}-D_{-m}D^{\dagger}_{-m})+
\delta_{m+\hat{a},0}(B^{\dagger}_{m}D^{\dagger}_{-m}+D_{-m}B_{m})\Big\}.
 \end{eqnarray*}
The last expression uses the identity $\Theta(m+a)-\Theta(-m-a)=sign(m+a)$ where ``$sign$'' is the sign function whose value
in $0$ is taken to be $sign(0)=0$. Furthermore the notation $\delta_{m+\hat{a},0}$ stands for a generalized Kronecker symbol
of which the indices may take continuous values, such that its value vanishes unless the two indices are equal in which case
the symbol takes the value unity. Once again the normal ordered form for the regularized operator $Q_{5}$ needs to be considered.
The Poisson resummation formula allows to isolate divergent contributions in \eref{somme1} and \eref{somme2}, leading to,
\begin{eqnarray*}
\sum_{m=-\infty}^{+\infty}(\Theta(m+a)-\Theta(-m-a))\exp[-\tilde{\alpha}(m+a)^2]\\
\stackrel{\tilde{\alpha}\rightarrow 0}{=} \frac{1}{2}(\sqrt{\frac{\pi}{\tilde{\alpha}}}-\sqrt{\frac{\pi}{\tilde{\alpha}}})+
\sum_{n=-\infty,n\ne 0}^{+\infty}\frac{\exp(2\rmi\pi na)}{\rmi\pi n}.
\end{eqnarray*}
Furthermore the series corresponds to the following Fourier expansion, provided $a$ is non integer (see the Appendix),
\begin{eqnarray}
\sum_{n=-\infty,n\ne 0}^{+\infty}\frac{\exp(2\rmi\pi na)}{\rmi\pi n}=1-2(a-\lfloor a\rfloor ).
\label{adelta}
\end{eqnarray}
However one needs to specify what the \emph{r.h.s.} of \eref{adelta} means when $a$ is an integer.
If the series in the \emph{l.h.s.} of \eref{adelta} is summed symmetrically, its value vanishes. Hence for the sake of
consistency, the final and complete expression for \eref{adelta} reads,
 \begin{eqnarray*}
\sum_{n=-\infty,n\ne 0}^{+\infty}\frac{\exp(2\rmi\pi na)}{\rmi\pi n}=1-2(a-\lfloor a\rfloor+\frac{1}{2}I(\hat{a})),
\end{eqnarray*}
where $I(a)$ stands for the discontinuous function which vanishes for all real values of $a$ except when $a$ is an integer,
$a\in\mathbb{Z}$, in which case $I(a)$ takes the value unity. It is also useful to keep in mind the property
$\lfloor -a\rfloor=-\lfloor a\rfloor-1 +I(a)$. Thus finally the fully gauge invariant expression of the axial charge,
which remains now well defined in the absence of a regulator, is
\clearpage
\begin{eqnarray}
: \hat{Q}_{5}:_{\hat{a}}=2(\hat{a}-\lfloor \hat{a}\rfloor )-1+I(\hat{a})\nonumber\\
+\sum_{m=-\infty}^{+\infty}\big[sign(m+\hat{a})(B^{\dagger}_{m}B_{m}+D^{\dagger}_{-m}D_{-m})+
\delta_{m+\hat{a},0}(B^{\dagger}_{m}D^{\dagger}_{-m}+D_{-m}B_{m})\big].\nonumber
\end{eqnarray}
This operator indeed generates global axial $U(1)_A$ transformations, for $m+\hat{a}\neq 0$,
\begin{eqnarray*}
B^{\dagger}_m &\rightarrow \exp(\rmi \beta)B^{\dagger}_m,\qquad
D^{\dagger}_{-m} &\rightarrow \exp(\rmi \beta)D^{\dagger}_{-m}\nonumber
\end{eqnarray*}
(for $m+\hat{a}=0$ an additional contribution arises because of the spectral flow properties in $\hat{a}$ of these operators).

Finally, let us point out that even though the Coulomb interaction contribution to the first class Hamiltonian has not been
considered explicitly so far, the reason for this is that a simple consideration of the expression \eref{eq:HC} for that operator $\hat{H}_C$
in terms of the fermionic modes readily shows that in the given form, it does not suffer quantum ordering ambiguities nor divergences since
no contribution with $\ell=0$ is involved in either of the two factors being multiplied in the sum over $\ell$.

\section{Modular Invariant Quantum Operators and the Axial Anomaly}

All potential divergences in the operators of interest having been subtracted consistently and in a manifestly modular invariant
manner, let us first now focus our attention on the global symmetry of the massless classical theory, namely its axial symmetry.
As is well-known these transformations are no longer a symmetry of the quantized dynamics because of a mechanism
that involves the ``topological'' zero-mode sector which, in the present formulation, is clearly identified. The gauge invariant
composite operators having been constructed so far include (the Casimir vacuum energy $(-\pi/(6L))$ is henceforth ignored in the
total first class Hamiltonian),
\begin{eqnarray}
:\hat{H}:_{\hat{a}} = \frac{(\hat{p}^{1})^2}{2L}+\frac{2\pi}{L}(\hat{a}-\lfloor \hat{a}\rfloor -\frac{1}{2})^2 \nonumber\\
+ \frac{2\pi}{L}\sum_{m}|m+\hat{a}|(B^{\dagger}_{m}B_{m}+D^{\dagger}_{-m}D_{-m}) + :\hat{H}_C:_{\hat{a}}, \label{quantumH}\\
:\hat{Q}_{5}:_{\hat{a}}=2(\hat{a}-\lfloor \hat{a}\rfloor -\frac{1}{2})+I(\hat{a}) +q_{5},\\
:\hat{Q}:_{\hat{a}}=\sum_{m}(B^{\dagger}_{m}B_{m}-D^{\dagger}_{-m}D_{-m}) \label{Qhat},
\end{eqnarray}
where 
\begin{eqnarray}
q_{5}=\sum_{m}\big[&sign(m+\hat{a})(B^{\dagger}_{m}B_{m}+D^{\dagger}_{-m}D_{-m})\nonumber\\
&+\delta_{m+\hat{a},0}(B^{\dagger}_{m}D^{\dagger}_{-m}+D_{-m}B_{m})\big],\label{q5}
\end{eqnarray}
while the gauge invariant total momentum operator of the system may be shown to be given as,
\begin{eqnarray}
:\hat{P}:_{\hat{a}}=\sum_{m}\frac{2\pi}{L}(m+\hat{a})(B^{\dagger}_{m}B_{m}-D^{\dagger}_{-m}D_{-m}).
\end{eqnarray}

Since the $B$ and $D$ operators and their adjoints depend on the operator $\hat{a}_1$ through the operator 
$\hat{a}=\frac{e\hat{a}_{1}L}{2\pi}-\lambda$,
the $B$ and $D$'s do not commute with the conjugate momentum of $\hat{a}_1$, namely $\hat{p}^{1}$. A direct
calculation finds,
\begin{eqnarray}
&[\hat{p}^{1},B^{\dagger}_{m}]=-\rmi\frac{eL}{2\pi}\delta(m+\hat{a})D_{-m},\qquad
&[\hat{p}^{1},D_{-m}]=\rmi\frac{eL}{2\pi}\delta(m+\hat{a})B^{\dagger}_{m}\label{pD},
\end{eqnarray}
as well as the corresponding adjoint relations (here, $\delta(m+\hat{a})$ stands for the usual Dirac $\delta$ function).
These results use the definitions \eref{2un} and \eref{2deux}
and the identity between distributions, $\partial_{x}\sqrt{\Theta(x)}=\delta(x)/\sqrt{2}$, given the choice $\Theta(0)=1/2$.
From these commutation relations it easily follows that $:\hat{H}:_{\hat{a}}$ commutes with $:\hat{Q}:_{\hat{a}}$.
However, the same is not true for the axial charge operator for which the calculation requires the evaluation of the
commutator $\big[\hat{p}^{1},:\hat{Q}_{5}:_{\hat{a}}\big]$. By differentiation of \eref{adelta}
and making use of \eref{pD}, one finds,
\begin{eqnarray}
:\big[\hat{p}^{1},:\hat{Q}_{5}:_{\hat{a}}\big]:_{\hat{a}}=-2\rmi\frac{eL}{2\pi},
\end{eqnarray}
and in turn finally, 
\begin{eqnarray}
:\big[:\hat{H}:_{\hat{a}},:\hat{Q}_{5}:_{\hat{a}}\big]:_{\hat{a}}=\
:\big[\frac{(\hat{p}^{1})^2}{2L},:\hat{Q}_{5}:_{\hat{a}}\big]:_{\hat{a}}=-2\rmi\frac{e\hat{p}^{1}}{2\pi}.
\end{eqnarray}
Since this relation expresses the quantum equation of motion for the axial charge in the Heisenberg picture, one observes
that this charge is no longer conserved, hence suffers a ``quantum anomaly''.
It is noticeable that this anomaly finds its origin only in the topological sector $(\hat{a}_{1},\hat{p}^{1})$.
The physical interpretation and consequences of this result have been discussed in the literature \cite{ColemanSusskind,Coleman1}.

\section{Modular Invariant Bosonization}

Rather than wanting to diagonalize the gauge invariant Hamiltonian for physical states,
it is possible to show that the theory describes in fact the dynamics of a free massive (pseudo)scalar boson of mass $m>0$
on the physical space, in the form,
\begin{eqnarray*}
:\hat{H}:_{\hat{a}}&=\frac{1}{2}:\Pi(0)^{\dagger}\Pi(0):_{\hat{a}}+\frac{1}{2}m^2:\Phi^{\dagger}(0)\Phi(0):_{\hat{a}}+\\
&+\frac{1}{2}\sum_{k\neq 0} :\Big\{\Pi^{\dagger}(k)\Pi(k)+(m^2+(\frac{2\pi k}{L})^{2})\Phi^{\dagger}(k)\Phi(k)\Big\}:_{\hat{a}}.
\end{eqnarray*}
The normal ordering prescription, $: \ :_{\hat{a}}$, for the fields $(\Phi(k),\Pi(k))$ will be specified hereafter.
As usual the scalar bosonic theory is defined by
\begin{eqnarray*}
H=\int_{S^{1}}dx\frac{1}{2}\big\{ \Pi^{\dagger}(x)\Pi(x)+\Phi^{\dagger}(x)(-\partial_{1}^{2}+m^{2})\Phi(x)\big\},
\end{eqnarray*}
with $\Phi(x)=1/\sqrt{L} \sum_{k}\Phi(k)e^{\rmi\frac{2\pi k x}{L}}$ and $\Pi(x)=1/\sqrt{L}\sum_{k}\Pi(k)e^{\rmi\frac{2\pi k x}{L}}$,
$\Pi(x)$ being the momentum canonically conjugate to $\Phi(x)$ and $k\in\mathbb{Z}$.

Let us now define the Fourier $k$-modes ($k\neq0$) for the boson and its conjugate momentum in terms of the fermionic modes as \cite{LZ,Manton:1985jm},
\clearpage
\begin{eqnarray*}
&\Phi(k)=\frac{-1}{\sqrt{2}\rmi k}\sqrt{\frac{L}{2\pi}}:(j_{1}(k)+j_{2}(k)):_{\hat{a}}\\
&\Pi(k)=\frac{1}{\sqrt{2}}\sqrt{\frac{2\pi}{L}}:(j_{1}(k)-j_{2}(k)):_{\hat{a}},
\end{eqnarray*}
where $j_{1}(k)=\sum_{m}b^{\dagger}_{m+k}b_{m}$ and $j_{2}(k)=\sum_{m}d_{-(m+k)}d^{\dagger}_{-m}$. Note that for $k\ne 0$
these operators are involved in the contributions to the Coulomb interaction energy.

These definitions ensure that the $k$-modes $\Phi(k)$ and $\Pi(k)$ fulfil the following necessary properties,
$\Phi^{\dagger}(k)=\Phi(-k)$ and $\Pi^{\dagger}(k)=\Pi(-k)$. For $k\ne 0$ the operators $j_{1}(k)$ and $j_{2}(k)$
may be expressed in terms of the $B$ and $D$ operators and their adjoints. Actually normal ordering of $j_j(k)$ ($j=1,2$) is only required
for $k=0$. As long as $k\ne 0$, no ordering ambiguity arises. By extension of the ordering procedure described in the previous Sections,
henceforth the normal ordered form, denoted $:\hat{O}:_{\hat{a}}$, of an operator $\hat{O}$ made of a product of $b^{(\dagger)}$'s
and $d^{(\dagger)}$'s is given by the normal ordered form with respect to the $B^{(\dagger)}$
and $D^{(\dagger)}$ operators upon the appropriate substitutions. However since intermediate steps in calculations or
partial contributions to quantities may produce divergent quantities, it should be wise to regularize expressions
before performing computations.

It being understood that the operators $j_j(k)$ are defined as has just been described, namely $j_j(k)\equiv:j_j(k):_{\hat{a}}$,
an explicit evaluation finds that these operators obey the following closed algebra,
\begin{eqnarray}
&:[j_{1}(k),j_{1}(\ell)]:_{\hat{a}}=\ell \delta_{k+\ell,0},\label{hier1}\\
&:[j_{2}(k),j_{2}(\ell)]:_{\hat{a}}=-\ell \delta_{k+\ell,0},\label{hier2}\\
&:[j_{1}(k),j_{2}(\ell)]:_{\hat{a}}=0.\label{hier3}
\end{eqnarray}
Let us establish here the first of these results. To compute the commutator \eref{hier1} consider the case when $k$ and $\ell$
have opposite signs (if they have the same sign it is easy to prove that the commutator vanishes), and introduce
the gaussian regularization procedure to handle potential divergences,
\begin{eqnarray}
[j_{1}(k),j_{1}(-\ell)]\nonumber\\
=\big[\sum_{m}b^{\dagger}_{m+k}b_{m}\exp(-\tilde{\alpha}(m+\hat{a})^2)\,,\,
\sum_{n}b^{\dagger}_{n}b_{n+\ell}\exp(-\tilde{\alpha}(n+\hat{a})^2)\big],\label{commut}
\end{eqnarray}
for $k,\ell>0$. Using the anti-commutation relations, in normal ordered form \eref{commut} becomes,
\begin{eqnarray*}
\sum_{m,n}&(:b^{\dagger}_{m+k}b_{n+\ell}:_{\hat{a}}\delta_{m,n}-:b^{\dagger}_{m}b_{n}:_{\hat{a}}\delta_{m+k,n+\ell})\\
&\times\exp[-\tilde{\alpha}(m+\hat{a})^2]\exp[-\tilde{\alpha}(n+\hat{a})^2].
\end{eqnarray*}
When substituted in terms of the $B$, $D$ operators, in the limit $\tilde{\alpha} \rightarrow 0$ this last expression reduces to,
\begin{eqnarray*}
\sum_{n}\exp[-2\tilde{\alpha}(n+\hat{a})^2](\Theta(-n-k-\hat{a})-\Theta(-n-\hat{a}))\delta_{k,\ell}
\stackrel{\tilde{\alpha}\rightarrow 0}{=}-k\delta_{k,\ell},
\end{eqnarray*}
which is indeed the result in \eref{hier1}. And from the commutation relations \eref{hier1} to \eref{hier3},
it readily follows that bosonic $k$-modes $(\Phi(k),\Pi(k))$ ($k\ne 0$) do indeed obey the Heisenberg algebra as it should,
\begin{eqnarray}
[\Phi(k),\Pi(\ell)]=\rmi\delta_{k+\ell,0},\qquad k,\ell\ne 0.
\end{eqnarray}

Let us now tackle the bosonized version of the Hamiltonian, by showing that it indeed
reproduces the expression \eref{quantumH}. The $k$-mode part of the bosonic Hamiltonian is
\begin{eqnarray*}
\frac{1}{2}\sum_{k\neq 0} :\big\{\Pi^{\dagger}(k)\Pi(k)+(\frac{2\pi k}{L})^{2}\Phi^{\dagger}(k)\Phi(k)\big\}:_{\hat{a}}\\
=\frac{1}{2}\frac{2\pi}{L}\sum_{k\neq 0}: \left(j_{1}^{\dagger}(k)j_{1}(k)+j_{2}^{\dagger}(k)j_{2}(k)\right):_{\hat{a}}.
\end{eqnarray*}
Using the commutation relations \eref{hier1} and \eref{hier2} one finds,
\begin{eqnarray}
&\frac{1}{2}\sum_{k\neq 0} :\{ \Pi^{\dagger}(k)\Pi(k)+(\frac{2\pi k}{L})^{2}\Phi^{\dagger}(k)\Phi(k)\}:_{\hat{a}}\nonumber\\
&=\frac{2\pi}{L}\sum_{k> 0} \sum_{m,n} 
:\{b^{\dagger}_{m+k}b_{m}b^{\dagger}_{n-k}b_{n}+d_{-(m-k)}d^{\dagger}_{-m}d_{-(n+k)}d^{\dagger}_{-n}\}:_{\hat{a}}\nonumber\\
&=\frac{2\pi}{L}\sum_{k> 0} \sum_{m,n}
:\{b^{\dagger}_{m+k}b_{m}b^{\dagger}_{n}b_{n+k}+d^{\dagger}_{-(m+k)}d_{-m}d^{\dagger}_{-n}d_{-(n+k)}\}:_{\hat{a}}.\label{ici}
\end{eqnarray}
A little algebra shows that the sum over the range of values
when $m\neq n$ vanishes on account of the anticommutation properties of the $b^{(\dagger)}_m$ and $d^{(\dagger)}_m$ operators.
Only the diagonal $m=n$ terms remain and provide the normal ordered expression,
\begin{eqnarray}
\sum_{k> 0} \sum_{m} :\left(b^{\dagger}_{m+k}b_{m}b^{\dagger}_{m}b_{m+k}
+d^{\dagger}_{-(m+k)}d_{-m}d^{\dagger}_{-m}d_{-(m+k)}\right):_{\hat{a}}.\label{la}
\end{eqnarray}
Substituting now for the $B^{(\dagger)}_m$ and $D^{(\dagger)}_m$ operators and using their anticommutation relations,
\eref{la} becomes in an explicitly normal ordered form,
\begin{eqnarray}
\sum_{k>0} \sum_{m}\Big[
&(B^{\dagger}_{m+k}B_{m+k}D^{\dagger}_{-m}D_{-m}+D^{\dagger}_{-(m+k)}D_{-(m+k)}B^{\dagger}_{m}B_{m}\big)\times\nonumber\\
&\times\Big(\Theta(m+k+\hat{a})\Theta(-m-\hat{a})\Big)\label{un}\\
&-\big(B^{\dagger}_{m+k}B_{m+k}B^{\dagger}_{m}B_{m}+D^{\dagger}_{-(m+k)}D_{-(m+k)}D^{\dagger}_{-m}D_{-m}\big)\times\nonumber\\
&\times\Big(\Theta(m+k+\hat{a})\Theta(m+\hat{a})+\Theta(-m-k-\hat{a})\Theta(-m-\hat{a})\Big)\label{deuxtrois}\\
&-(B^{\dagger}_{m+k}B_{m+k}+D^{\dagger}_{-(m+k)}D_{-(m+k)})(B^{\dagger}_{m}D^{\dagger}_{-m}+D_{-m}B_{m})\times\nonumber\\
&\times\frac{1}{2}\Theta(m+\hat{a}+k)\delta_{m+\hat{a},0}\label{troisbis}\\
&+(B^{\dagger}_{m}B_{m}+D^{\dagger}_{-m}D_{-m})(B^{\dagger}_{m+k}D^{\dagger}_{-(m+k)}+D_{-(m+k)}B_{m+k})\times\nonumber\\
&\times\frac{1}{2}\Theta(-m-\hat{a})\delta_{m+k+\hat{a},0}\label{troistris}\\
&+(B^{\dagger}_{m+k}B_{m+k}+D^{\dagger}_{-(m+k)}D_{-(m+k)})\Theta(m+k+\hat{a})\Theta(m+\hat{a}) \label{quatre}\\
&+(B^{\dagger}_{m}B_{m}+D^{\dagger}_{-m}D_{-m})\Theta(-m-k-\hat{a})\Theta(-m-\hat{a})\Big]\label{cinq}.
\end{eqnarray}
The first eight lines \eref{un} to \eref{troistris} are quadrilinear in the $B^{(\dagger)}$ and $D^{(\dagger)}$ operators while the last two
lines \eref{quatre} and \eref{cinq} are bilinear. They need to be handled differently.

The quadrilinear terms combine to give
\begin{eqnarray}
-\frac{1}{4}(\hat{Q}^2+q_{5}^2)+\frac{1}{2}\Big[\sum_{m}(1-\delta_{m+\hat{a},0})(B^{\dagger}_{m}B_{m}+D^{\dagger}_{-m}D_{-m})\Big]
+\frac{1}{4}I(\hat{a}),\label{quadri}
\end{eqnarray}
with the help of \eref{Qhat} and \eref{q5}, as may be checked by writing out \eref{quadri} explicitly.

The bilinear terms in \eref{quatre} and \eref{cinq} may be written as,
\begin{eqnarray}
{\ }\hspace{-10pt}
&\sum_{k>0}\sum_{m} [N_{m+k}\Theta(m+k+\hat{a})\Theta(m+\hat{a})+N_{m}\Theta(-m-k-\hat{a})\Theta(-m-\hat{a})]\nonumber\\
{\ }\hspace{-10pt}
&=\sum_{m}N_{m}\left\{\sum_{k>0}[\Theta(m+\hat{a})\Theta(m+\hat{a}-k)+\Theta(-m-\hat{a})\Theta(-m-\hat{a}-k)]\right\},\nonumber
\label{eq:2lin}
\end{eqnarray}
where $N_{m}=(B^{\dagger}_{m}B_{m}+D^{\dagger}_{-m}D_{-m})$.
Let us focus on any one of the terms in the series in curly brackets for any specific value of $m\in\mathbb{Z}$,
in which $N_m$ is multiplied by the following series,
\begin{eqnarray}
\sum_{k>0}[\Theta(m+a)\Theta(m+a-k)+\Theta(-m-a)\Theta(-m-a-k)].\label{term}
\end{eqnarray}
If $m+a=0$ this latter quantity vanishes explicitly since $\Theta(-k)=0$ for $k>0$.
Consider then the case when $m+a\neq0$. Making use of the identity
\begin{eqnarray}
\sum_{k=1}^{+\infty}\theta(x-k)=\lfloor x \rfloor -\frac{1}{2}I(x),
\end{eqnarray}
which applies only for $x>0$, one finds,
\begin{eqnarray}
&\Theta(m+a)\sum_{k>0}\Theta(m+a-k)=\Theta(m+a)\left(\lfloor m+a \rfloor -\frac{1}{2}I(a)\right),\nonumber\\
&\Theta(-m-a)\sum_{k>0}\Theta(-m-a-k)=\Theta(-m-a)\left(\lfloor -m-a \rfloor -\frac{1}{2}I(a)\right).\nonumber
\end{eqnarray}
However since one has,
\begin{eqnarray}
&\lfloor m+a \rfloor=m+\lfloor a \rfloor,\qquad
\lfloor -(m+a) \rfloor=-\lfloor m+a \rfloor-1+I(a),\nonumber
\end{eqnarray}
the series \eref{term} takes the form,
\begin{eqnarray*}
\Theta(m+a)\left( m+\lfloor a \rfloor -\frac{1}{2}I(a)\right)\\
+\Theta(-m-a)\left(- m-\lfloor a \rfloor-1+I(a) -\frac{1}{2}I(a)]\right),
\end{eqnarray*}
or equivalently,
\begin{eqnarray}
&\Theta(m+a)\left( m+a -a+\lfloor a \rfloor -\frac{1}{2}I(a)\right)\\
&+\Theta(-m-a)\left( - m-a+a-\lfloor a \rfloor+\frac{1}{2}I(a)\right)-\theta(-m-a).
\end{eqnarray}
Using now the fact that $\Theta(-m-a)=(1-sign(m+a))/2$ the series \eref{term} finally takes the following expression when
$m+a\ne 0$,
\begin{eqnarray}
&|m+a|-sign(m+a)\left(a-\lfloor a \rfloor+\frac{1}{2}I(a)\right)-\frac{1}{2}\left(1-sign(m+a)\right)\nonumber\\
&=|m+a|-\frac{1}{2}-sign(m+a)\left(a-\lfloor a \rfloor-\frac{1}{2}+\frac{1}{2}I(a)\right).
\end{eqnarray}
Since the series \eref{term} vanishes when $m+a=0$, the complete expression may be written by subtracting from the above result
its value when $m+a=0$, producing the final expression for the series \eref{term},
\begin{eqnarray}
|m+a|-\frac{1}{2}-sign(m+a)\left(a-\lfloor a \rfloor-\frac{1}{2}+\frac{1}{2}I(a)\right)+\frac{1}{2}\delta_{m+a,0},
\end{eqnarray}
valid for any $m\in \mathbb{Z}$ and any $a\in\mathbb{R}$.

Substituting this identity in \eref{eq:2lin}, one finally obtains for the sum of \eref{quatre} and \eref{cinq},
\begin{eqnarray}
\frac{1}{2}\sum_{m}\delta_{m+\hat{a},0}N_{m}
+\sum_{m}(|m+\hat{a}|-\frac{1}{2})N_{m}\nonumber\\
-(\hat{a}-\lfloor \hat{a}\rfloor -\frac{1}{2}+\frac{1}{2}I(\hat{a}))\sum_{m}sign(m+\hat{a})N_{m}.
\label{bilin}
\end{eqnarray}
Then the sum of \eref{bilin} and \eref{quadri} leads to the following expression for the $k$-mode contribution ($k\ne 0$) to the
bosonic Hamiltonian,
\begin{eqnarray}
\frac{2\pi}{L}\Big(\sum_{m}|m+\hat{a}|N_{m}-(\hat{a}-\lfloor \hat{a}\rfloor -\frac{1}{2}+\frac{1}{2}I(\hat{a}))
\sum_{m}sign(m+\hat{a})N_{m}\nonumber\\
-\frac{1}{4}(\hat{Q}^2+q_{5}^2)+\frac{1}{4}I(\hat{a})\Big).\label{inter}
\end{eqnarray}
Obviously this last expression includes the fermionic bilinear contribution to the Hamiltonian in \eref{quantumH}.
Furthermore \eref{inter} gives also a clue for the zero-mode part of the bosonized Hamiltonian.
Let us complete a square as follows,
\begin{eqnarray}
\frac{2\pi}{L}\Big(\sum_{m}|m+\hat{a}|N_{m}-\Big(\hat{a}-\lfloor \hat{a}\rfloor -
\frac{1}{2}+\frac{1}{2}I(\hat{a})+\frac{1}{2}q_{5}\Big)^2\nonumber\\
+\big(\hat{a}-\lfloor \hat{a}\rfloor -\frac{1}{2}+\frac{1}{2}I(\hat{a})\big)^2-\frac{1}{4}\hat{Q}^2+\frac{1}{4}I(\hat{a})\Big),\label{square}
\end{eqnarray}
with $q_{5}$ given in \eref{q5} and where the contribution in $\hat{Q}^2$ vanishes for the physical states.
Indeed this last relation applies since one has the property 
\begin{eqnarray*}
(a-\lfloor a\rfloor -\frac{1}{2}+\frac{1}{2}I(\hat{a}))q_{5}\\
=(a-\lfloor a\rfloor -\frac{1}{2}+\frac{1}{2}I(\hat{a}))\sum_{m}sign(m+a)N_{m},
\end{eqnarray*}
given the expression in \eref{adelta} and the fact that the product of $\delta_{m+\hat{a},0}$ with the first factor in this last expression 
vanishes identically. Likewise by direct expansion, one finds,
\clearpage
\begin{eqnarray*}
\big(\hat{a}-\lfloor \hat{a}\rfloor -\frac{1}{2}+\frac{1}{2}I(\hat{a})\big)^2+\frac{1}{4}I(\hat{a})\\
=\big(\hat{a}-\lfloor \hat{a}\rfloor - \frac{1}{2}\big)^2+\big(\hat{a}-\lfloor \hat{a}\rfloor - \frac{1}{2}\big)I(\hat{a})
+\frac{1}{4}I(\hat{a})+\frac{1}{4}I(\hat{a})\\
=\big(\hat{a}-\lfloor \hat{a}\rfloor - \frac{1}{2}\big)^2.
\end{eqnarray*}

We may now complete the bosonization procedure and define the missing pieces in the bosonized formulation.
One needs to identify the bosonic conjugate momentum zero-mode, $\Pi(0)$. The result \eref{square} provides this identification through,
\begin{eqnarray*}
\frac{1}{2}\Pi(0)^{\dagger}\Pi(0)=\frac{2\pi}{L}\Big(\hat{a}-\lfloor \hat{a}\rfloor
-\frac{1}{2}+\frac{1}{2}I(\hat{a})+\frac{1}{2}q_{5}\Big)^2=\frac{\pi}{2L}\left(:\hat{Q}_5:_{\hat{a}}\right)^2,
\end{eqnarray*}
hence one defines,
\begin{eqnarray}
\Pi(0)=\pm\sqrt{\frac{\pi}{L}}:\hat{Q}_{5}:_{\hat{a}}.\label{pm}
\end{eqnarray}
To sum up we have established the following identity, which is valid for physical states only with $\hat{Q}=0$,
\begin{eqnarray*}
&\frac{1}{2}:\Pi(0)^{\dagger}\Pi(0):_{\hat{a}}
+\frac{1}{2}\sum_{k\neq 0}:\left(\Pi^{\dagger}(k)\Pi(k)+(\frac{2\pi k}{L})^{2}\Phi^{\dagger}(k)\Phi(k)\right):_{\hat{a}}\\
&=\frac{2\pi}{L}\big(\hat{a}-\lfloor \hat{a}\rfloor -\frac{1}{2}\big)^2+\frac{2\pi}{L}
\sum_{m}|m+\hat{a}|(B^{\dagger}_{m}B_{m}+D^{\dagger}_{-m}D_{-m}).
\end{eqnarray*}
Finally the Coulomb interaction Hamiltonian provides the mass term for the boson,
\begin{eqnarray*}
\frac{1}{2}\sum_{k\neq 0} m^{2}:\Phi(k)^{\dagger}\Phi(k):_{\hat{a}}\nonumber\\
=\frac{e^{2}L}{2(2\pi)^{2}}\sum_{k\neq 0}:\frac{(:j_{1}^{\dagger}(k):_{\hat{a}}+:j_{2}^{\dagger}(k):_{\hat{a}})
(:j_{1}(k):_{\hat{a}}+:j_{2}(k):_{\hat{a}})}{k^2}:_{\hat{a}},
\end{eqnarray*}
hence the identification $m^2=e^2/\pi$.
And the very last piece of the puzzle is the zero-mode of the boson, $\Phi(0)$, provided by,
\begin{eqnarray}
\frac{1}{2}m^2\Phi^{\dagger}(0)\Phi(0)=\frac{(\hat{p}^{1})^2}{2L},\nonumber
\end{eqnarray}
which leads to $\Phi(0)=\sqrt{\pi}\hat{p}^{1}/(e\sqrt{L})$. The choice of sign for this quantity is correlated to that
of the conjugate momentum zero mode, $\Pi(0)$. By choosing the minus sign for the square root in \eref{pm},
one then also obtains the proper Heisenberg algebra for the boson zero-modes,
\begin{eqnarray}
:\big[\Phi(0)\, , \,\Pi(0)\big]:_{\hat{a}}=\frac{\pi}{eL}:\big[\hat{p}^{1}\, ,\, -:\hat{Q}_{5}:_{\hat{a}}\big]:_{\hat{a}}=\rmi.
\end{eqnarray}
The axial anomaly thus proves to be central in establishing the correct commutation relation in the zero-mode sector of the
bosonized fermion.

In conclusion, when restricted to the space of physical quantum states of total vanishing electric charge, $\hat{Q}=0$,
the total first class Hamiltonian, whether expressed in terms of the original fermion modes or the bosonic ones given by
\begin{eqnarray*}
\Phi(0)&=\sqrt{\pi}\frac{\hat{p}^{1}}{e\sqrt{L}},\qquad
&\Phi(k\ne 0)=\frac{-1}{\sqrt{2}\rmi k}\sqrt{\frac{L}{2\pi}}:(j_{1}(k)+j_{2}(k)):_{\hat{a}},\\
\Pi(0)&=\pm\sqrt{\frac{\pi}{L}}:\hat{Q}_{5}:_{\hat{a}},\qquad
&\Pi(k\ne 0)=\frac{1}{\sqrt{2}}\sqrt{\frac{2\pi}{L}}:(j_{1}(k)-j_{2}(k)):_{\hat{a}},
\end{eqnarray*}
determines the same quantum theory and physical content.

\section{Adding a Theta Term}

A natural extension of this low dimensional model is the inclusion of a ``theta'' term, which is the analogue of the topological
$\theta$ term in four dimensional QCD, by adding the following contribution to the original Lagrangian density of the Schwinger model,
\begin{eqnarray}
\mathcal{L}_{\theta}=\frac{e}{2}\frac{\theta}{2\pi}\epsilon_{\mu \nu}F^{\mu\nu},
\end{eqnarray}
where the parameter $\theta$ has a dimension of mass. The entire analysis of constraints can be carried through once again in a manner
similar to what has been done previously, leading to the following first class quantum Hamiltonian corresponding to the one
in \eref{quantumH},
\begin{eqnarray}
:\hat{H}:_{\hat{a}} &= \frac{1}{2L}(\hat{p}^1- eL\frac{\theta}{2\pi})^{2}
+\frac{2\pi}{L}(\hat{a}-\lfloor \hat{a}\rfloor -\frac{1}{2})^2\nonumber\\
 &+ \sum_{m}\frac{2\pi}{L}|m+\hat{a}|(B^{\dagger}_{m}B_{m}+D^{\dagger}_{-m}D_{-m}) +:\hat{H}_C:_{\hat{a}} .
\end{eqnarray}
The shift by a term proportional to $\theta$ in the contribution of the gauge zero-mode conjugate momentum $\hat{p}^1$
is also observed in the axial anomaly,
\begin{eqnarray}
:\big[:\hat{H}:_{\hat{a}}\,,\, :\hat{Q}_{5}:_{\hat{a}}\big]:_{\hat{a}}
&=:\big[\frac{(\hat{p}^{1}-eL\theta/2\pi)^2}{2L}\, , \, :\hat{Q}_{5}:_{\hat{a}}\big]:_{\hat{a}}\\
&=-\rmi\frac{e^{2}}{\pi}L\left(\frac{\hat{p}^{1}}{eL}-\theta/2\pi\right).
\end{eqnarray}

Given this observation which applies to the model with a massless fermion, it should be clear that all previous
considerations remain valid in terms of the shifted conjugate momentum, $(\hat{p}^1-eL\theta/(2\pi))$, which still defines
a Heisenberg algebra with the gauge zero mode $\hat{a}_1$. Note that the introduction of the shifted variable affects the
modular transformation operators $\hat{U}(\ell)$ only by a redefinition of the arbitrary phase factor $\theta_0$
as $\theta_0\rightarrow \theta_0-\theta$, with no further consequence. Hence, in the massless fermion model,
the introduction of the $\theta$ term does not lead to a modified gauge invariant physical content of the quantized system.
It still is equivalent to a theory of a free (pseudo)scalar bosonic field of mass $m=|e|/\sqrt{\pi}>0$.

\section{Conclusions}

In order to better understand the relevance and physical consequences of the topological sectors of gauge invariant dynamics,
the present work developed a careful analysis of the Schwinger model in its fermionic formulation
on a compactified spacetime with the cylindrical topology, within a manifestly gauge invariant formulation
without resorting to any gauge fixing procedure.
Among different reasons for considering a spatial compactification, one feature proves to be central to the discussion,
namely that of large gauge or modular transformations which capture the topologically non trivial characteristics of the
dynamics. Through proper regularization a quantization that remains manifestly invariant under modular transformations
is feasible, and allows at the same time a clear separation between locally gauge variant and invariant degrees of freedom
and globally gauge variant and invariant degrees of freedom, the latter being acted on by modular transformations only.
Spatial compactification brings to the fore all the subtle aspects related to the topological sectors and their dynamics of the model.

What proves to be a most remarkable fact indeed, which remains relevant more generally for any non-abelian Yang-Mills theory coupled
to charged matter fields in higher spacetime dimensions as well, is that the topologically non trivial modular gauge transformations
act by mixing the small and large distance and energy scales of the dynamics, a feature which is intrinsically non-perturbative as well
and thus cannot be captured through any perturbation theory that includes gauge invariance under small gauge transformations only.

To the authors' best knowledge such an analysis of the Schwinger model has not been available in the literature so far.
Besides recovering the well known result that as soon as the gauge coupling constant of the electromagnetic interaction
is turned on this theory is in actual fact that of a free spin zero massive particle in two dimensions, rather than
a theory of electrons and positrons coupled to photons, the analysis provides an original insight into the role played
by topology and modular invariance in a mechanism leading to the confinement of charged particles in an abelian gauge theory.
The fact that the chiral anomaly also finds its sole origin in the purely topological gauge sector is clearly made
manifest through the considered separation of variables which is devoid of any gauge fixing procedure whatsoever.
And finally the bosonization of the massless fermion is done at the operator level in terms of the fermionic modes
rather than through vertex operators of the boson, by paying due care and attention to the contributions of the
topological sector which again are crucial for the quantum equivalence between the two theories. In particular a manifestly
modular invariant bosonization of the fermion degrees of freedom has been achieved.

Having developed this approach in the specific laboratory of the Schwinger model, it remains now to extend these
methods to higher dimensions and explore how they may shed new insight into the topological and non-perturbative
dynamics of gauge theories coupled to charged matter, and the physical consequences of modular transformations
leading to ultraviolet and infrared mixing of the quantum modes of the charged matter fields. Such work is now in progress
in the case of 2+1 dimensional quantum electrodynamics.

\vspace{-10pt}

\ack

Florian Payen is warmly thanked for discussions and his contributions to the early stages of the present work.
It is a pleasure to acknowledge Se\'an Murray for comments on properties of generalised functions, and
Mathieu Buchkremer for stimulating discussions. The work of MF is supported by the National Fund for Scientific Research
(F.R.S.-FNRS, Belgium) through a ``Aspirant'' Research fellowship. This work is supported by the Belgian Federal Office
for Scientific, Technical and Cultural Affairs through the Interuniversity Attraction Pole P6/11.

\appendix
\section{Divergences in the Charge Operators}

In order to extract finite contributions out of otherwise divergent quantities, some regularization procedure is required,
for which either a gaussian or a zeta function regularization has been considered. The details of either regularization
leading to the results quoted in the main text are discussed in this Appendix. For simplicity
calculations are developed hereafter when the real variable $a$ is non integer. Extending results to the case when
$a\in \mathbb{Z}$ is discussed in the main text where appropriate.

\subsection{Gaussian Regularization}

The Poisson resummation formula may be used to establish the relation,
\begin{eqnarray}
{\ }\hspace{-20pt}
\sum_{m=-\infty}^{+\infty}\Theta(m+a)\exp[-\alpha(m+a)^2]\ \stackrel{\alpha\rightarrow 0}{=} \ \frac{1}{2}\sqrt{\frac{\pi}{\alpha}}+
\sum_{n=-\infty,n\ne 0}^{+\infty}\frac{\exp(2\rmi\pi na)}{2\rmi\pi n},
\label{divbil}
\end{eqnarray}
so that the subtraction of the short distance divergence consists in removing the term in $(1/2)\sqrt{\pi/\alpha}$.

To prove this result, one applies the Poisson resummation formula to the expression on the \emph{l.h.s.} of this relation,
in terms of the function $f(x)=\Theta(x+a)\exp[-\alpha(x+a)^2]$ of which the Fourier transform is, where $k\in\mathbb{R}$,
\begin{eqnarray}
\tilde{f}(k)&=\int_{-\infty}^{+\infty}dx \exp(-\rmi kx)\Theta(x+a)\exp[-\alpha(x+a)^2]\nonumber\\
&=\exp(\rmi ka)\,I^0_\alpha(k),\nonumber
\end{eqnarray}
with the definition
\begin{eqnarray}
I^{0}_{\alpha}(k)=\int_{0}^{+\infty}dx \exp(-\rmi kx-\alpha x^2),\nonumber
 \end{eqnarray}
so that,
\begin{eqnarray}
\sum_{m=-\infty}^{+\infty}\Theta(m+a)\exp[-\alpha(m+a)^2]=
\sum_{n=-\infty}^{+\infty}\tilde{f}(2\pi n).
\end{eqnarray}
Quite obviously $I^0_\alpha(0)=\frac{1}{2}\sqrt{\frac{\pi}{\alpha}}$, while for $k\ne 0$ the integral $I^0_\alpha(k)$ is expressed
in terms of the parabolic cylinder function $D_{-1}(z)$ \cite{Gradshteyn:1965},
\begin{eqnarray}
I^0_\alpha(k)=\frac{1}{\sqrt{2\alpha}}\,\exp\left(-\frac{k^2}{8\alpha}\right)\,D_{-1}\left(\frac{ik}{\sqrt{2\alpha}}\right).
\end{eqnarray}
Since the asymptotic behaviour of $D_{-1}(z)$ is known as $|z|\rightarrow+\infty$ \cite{Gradshteyn:1965},
in the small $\alpha$ limit one finds that $I^0_\alpha(k)$ behaves such that for $n\ne 0$,
\begin{eqnarray}
I^0_\alpha(2\pi n)\stackrel{\alpha\rightarrow 0}{=}\frac{1}{2\rmi\pi n}\left(1+{\cal O}(\alpha)\right).
\end{eqnarray}
Consequently, one has established the relation \eref{divbil}, with the further observation that the infinite series
contribution on the \emph{r.h.s.} is the Fourier series of a simple function of~$a$, when $a$ is non integer,
\begin{eqnarray}
\sum_{n=-\infty,n\ne 0}^{+\infty}\frac{\exp(2\rmi\pi na)}{2\rmi\pi n}=
\sum_{n=1}^{+\infty}\frac{\sin(2\pi na)}{\pi n}=\frac{1}{2}-(a-\lfloor a\rfloor ).
\end{eqnarray}

\subsection{Zeta Function Regularization}

A regularization of the $\zeta$ function type\footnote{This is also the regularization used in \cite{Manton:1985jm,Itoi:1991ch}.} of the same infinite series takes the following form, with $\alpha>0$
and in the limit $\alpha\rightarrow 0$,
 \begin{eqnarray*}
\sum_{m=-\infty}^{+\infty}\Theta(m+a)\exp[-\alpha(m+a)]=\exp\left(-\alpha(a+\lfloor -a\rfloor )\right)\left(\frac{1}{1-e^{-\alpha}}-1\right)\\
=\frac{1}{\alpha}-(a-\lfloor a\rfloor )+\frac{1}{2}+{\cal O}(\alpha),\nonumber
\end{eqnarray*}
when using $\lfloor -a\rfloor=-\lfloor a\rfloor-1$ (which applies when $a$ is non integer). 
Similarly,
\begin{eqnarray}
\sum_{m=-\infty}^{+\infty}\Theta(-m-a)\exp[\alpha(m+a)]=\frac{1}{\alpha}+(a-\lfloor a\rfloor )-\frac{1}{2}+{\cal O}(\alpha).\nonumber
\end{eqnarray}
Hence either regularization prescription produces the same finite contribution as a function of $a$
from the divergent series $\sum_{m=-\infty}^{+\infty}\Theta(m+a)$.

\section{Divergences in the Bilinear Fermion Hamiltonian}

\subsection{Gaussian Regularization}

We need also to show that
\begin{eqnarray}
\sum_{m=-\infty}^{+\infty}|m+a|\nonumber\\
 \stackrel{\alpha\rightarrow 0}{=}
 \ \sum_{m=-\infty}^{+\infty}(m+a)\big(\Theta(m+a)-\Theta(-m-a)\big)\exp[-\alpha(m+a)^2]\nonumber\\
=\frac{1}{\alpha}-2\sum_{n=-\infty,n\ne 0}^{+\infty}\frac{\exp(2\rmi\pi n a)}{(2\pi n)^2} + {\cal O}(\alpha).
\label{eq:Id3}
\end{eqnarray}
To make use of the Poisson resummation formula consider the function
 \begin{eqnarray}
g(x)=\big(\Theta(x+a)-\Theta(-x-a)\big)(x+a)\exp[-\alpha(x+a)^2],\nonumber
 \end{eqnarray}
of which the Fourier transform is, with $k\in\mathbb{R}$,
  \begin{eqnarray}
\tilde{g}(k)=\exp(\rmi ka)(I_{\alpha}(k)+I_{\alpha}(-k)),\nonumber
 \end{eqnarray}
where
 \begin{eqnarray}
I_{\alpha}(k)=\int_{0}^{+\infty}dx\ x\ \exp(-\rmi kx)\exp(-\alpha x^2),\nonumber
 \end{eqnarray}
whose value is expressed in terms of yet another parabolic cylinder function \cite{Gradshteyn:1965},
 \begin{eqnarray}
 I_{\alpha}(k)=\frac{1}{2\alpha}\Gamma(2) \exp(-\frac{k^2}{8\alpha})D_{-2}(\frac{\rmi k}{\sqrt{2\alpha}}).\nonumber
 \end{eqnarray}
Given the asymptotic behaviour of $D_{-2}(z)$ \cite{Gradshteyn:1965}, for $n\ne 0$ one finds in the limit $\alpha\rightarrow 0$,
\begin{eqnarray}
I_{\alpha}(n)\stackrel{\alpha\rightarrow 0}{=}-\frac{1}{n^2},\nonumber
\end{eqnarray}
while for $n=0$, $I_\alpha(0)=\frac{1}{2\alpha}$.
In conclusion, one has established that
\begin{eqnarray}
\sum_{m=-\infty}^{+\infty}g(m)=\sum_{n=-\infty}^{+\infty}\tilde{g}(2\pi n)\nonumber\\
\stackrel{\alpha\rightarrow 0}{=} \frac{1}{\alpha}-2\sum_{n=-\infty,n\ne 0}^{+\infty}
\frac{\exp(2\rmi\pi n a)}{(2\pi n)^2}= \frac{1}{\alpha}-(a-\lfloor a\rfloor -\frac{1}{2})^2+\frac{1}{12},
\label{divbis}
\end{eqnarray}
which is the relation in \eref{eq:Id3}.

\subsection{Zeta Function Regularization}

Using a $\zeta$ function regularization leads to the same result, namely,
\begin{eqnarray}
\sum_{m=-\infty}^{+\infty}|m+a|\nonumber\\
\stackrel{\alpha\rightarrow 0}{=} \ \sum_{m=-\infty}^{+\infty}((m+a)\Theta(m+a)e^{-\alpha(m+a)}-(m+a)\Theta(-m-a)e^{\alpha(m+a)})\nonumber\\
\stackrel{\alpha\rightarrow 0}{=} \ \frac{2}{\alpha^2}-(a-\lfloor a\rfloor -\frac{1}{2})^2+\frac{1}{12}.
\label{eq:Id4}
\end{eqnarray}
By defining
 \begin{eqnarray}
S_{+}=\sum_{m=-\infty}^{+\infty}(m+a)\Theta(m+a)\exp[-\alpha(m+a)],\nonumber
\end{eqnarray}
one observes that,
\begin{eqnarray}
S_{+}=-\frac{\partial}{\partial\alpha}\left(\exp[-\alpha(a+\lfloor -a\rfloor )](\frac{1}{1-e^{-\alpha}}-1)\right),\nonumber
\end{eqnarray}
of which a Laurent series expansion in $\alpha$ produces,
\begin{eqnarray}
S_{+}=\frac{1}{\alpha^2}-\frac{1}{2}(a-\lfloor a\rfloor -\frac{1}{2})^2+\frac{1}{24}.\nonumber
\end{eqnarray}
Similarly given
\begin{eqnarray}
S_{-}=\sum_{m=-\infty}^{+\infty}(m+a)\Theta(-m-a)\exp[\alpha(m+a)],\nonumber
\end{eqnarray}
this quantity takes the form
 \begin{eqnarray}
S_{-}=-\frac{1}{\alpha^2}+\frac{1}{2}(a-\lfloor a\rfloor -\frac{1}{2})^2-\frac{1}{24}=-S_{+}.\nonumber
\end{eqnarray}
Hence indeed the relation \eref{eq:Id4} has been established.

\end{document}